\documentclass[prb,twocolumn,showpacs,preprintnumbers,amsmath,amssymb]{revtex4-1}

\usepackage{graphicx}
\usepackage{ulem}
\usepackage{xcolor}

\begin{document}
\title{ Symmetry protected bosonic topological phase transitions: Quantum Anomalous Hall system of weakly interacting
       spinor bosons in a square lattice }
\author{ Fadi Sun$^{1,2,4}$, Junsen Wang $^{1,3}$, Jinwu Ye $^{1,2,4}$, Shaui Chen $^{3}$  and  Youjin Deng $^{3}$  }
\affiliation{
$^{1}$ Department of Physics and Astronomy, Mississippi State University, MS, 39762, USA \\
$^{2}$ Department of Physics, Capital Normal University,
Key Laboratory of Terahertz Optoelectronics, Ministry of Education, and Beijing Advanced innovation Center for Imaging Technology,
Beijing, 100048, China   \\
$^{3}$ CAS Center for Excellence and Synergetic Innovation Center in Quantum Information and Quantum Physics,
University of Science and Technology of China, Hefei, Anhui 230026, China  \\
$^{4}$ Kavli Institute of Theoretical Physics, University of California, Santa Barbara, Santa Barbara, CA 93106  }
\date{\today }

\begin{abstract}
 We study possible many body phenomena in the  Quantum Anomalous Hall system of weakly interacting spinor bosons in a square lattice.
 There are various novel spin-bond correlated superfluids (SF) and  quantum or topological  phase transitions among these SF phases.
 One transition is a first order one driven by  roton droppings ( but with non-zero gaps $ \Delta_R $ ) tuned by the Zeeman field $ h $.
 Another is a second order bosonic Lifshitz transition with the dynamic exponents $ z_x=z_y=2 $ and
 an accompanying $ [C_4 \times C_4]_D $ symmetry breaking. It is
 driven by the softening of the superfluid Goldstone mode tuned by the ratio of
 spin-orbit coupled (SOC) strength over the hopping strength.
 The two phase boundaries meet at a topological tri-critical (TT) point which separates the $ h=0 $ line into two SF phases
 with $ N=2 $ and $ N=4 $ condensation momenta respectively.
 At the $ h=0 $ line where the system has an anti-unitary  $ Z_2 $ Reflection symmetry, there are infinite number of
 classically degenerate family of states on both sides. We perform a systematic order from quantum disorder  analysis to find the
 quantum ground states, also calculate the roton gaps $ \Delta_R $ generated by the order from disorder mechanism on both sides of the TT point. The $ N=2 $ and $ N=4 $ SF phases have the same spin-orbital XY-AFM spin structure, respect the anti-unitary symmetry
 and break the $ [C_4 \times C_4]_D $ symmetry, so they be distinguished only by
 the different topology of the BEC condensation momenta instead of by any differences in the symmetry breaking patterns.
 This could be a first bosonic analog of the fermionic topological Lifshitz  transition with the change of the topology of
 the Fermi surfaces, Dirac points or Weyl points.
 However, when moving away from $ h=0 $ line, the $ Z_2 $ Reflection  symmetry is lost, the TT is converted to
 the second order bosonic Lifshitz transition.
 All these novel quantum or topological phenomena can be probed in the recent experimentally realized weakly
 interacting Quantum Anomalous Hall (QAH) model of $ ^87 Rb $ by Wu, {\sl et.al}, Science 354, 83-88 (2016).
\end{abstract}

\maketitle

The investigation and control of spin-orbit coupling (SOC) have become
subjects of intensive research in both condensed matter and cold atom systems after the discovery
of the topological insulators \cite{kane,zhang}.
In materials side,
the Rashba SOC plays crucial roles in various 2d or layered insulators, semi-conductor systems, metals and superconductors without inversion symmetry.
The Quantum Anomalous Hall ( QAH ) effect was experimentally realized
in Cr doped Bi(Sb)$_2$Te$_3$ thin films \cite{QAHthe,QAHexp} and also observed in many other materials such as both Cr doped and V doped (Bi,Sb)$_2$Te$_3$ films. In the cold atom side,
 using Raman scheme, one experimental group \cite{expk40,expk40zeeman}  generated 2d Rashba SOC
for $ ^{40} K $ gas. Using the optical Raman lattice scheme,
the bosonic analog of the QAH for spinor bosons $ ^{87}$Rb  was realized in \cite{2dsocbec}.
It was known that one serious shortcoming for using Raman scheme to generate SOC in alkali fermions,
is the strong heating associated with spontaneous emissions.
The heating issue hindered the observation of true many body phenomena for alkali fermions.
So the physics observed in  \cite{expk40,expk40zeeman} is still
at single particle physics. However, the heating issue is much less serious for spinor boson $ ^{87}$Rb  atoms in the weakly interacting regime.
Indeed, the lifetime of SOC $ ^{87}$Rb  BEC was already made as long as $ 300 ms $ in \cite{2dsocbec} and improved to
$ 1s $ recently \cite{more}, so the current experiment set-up the stage to observe any possible many body phenomena at a weak interaction
where the heating rate is well under current experimental control.
More recently,  optical lattice clock schemes \cite{SD,clock} have been successfully implemented \cite{clock1,clock2} to generate 1d SOC for $^{87} Sr $ and $ ^{137} Yb $. This newly developed scheme has the advantage to suppress
the heating issue suffered in the Raman scheme.
It can also be used to probe the interplay between the interactions and the SOC easily.

\begin{figure}[!htb]
\centering
    \includegraphics[width=0.45 \textwidth]{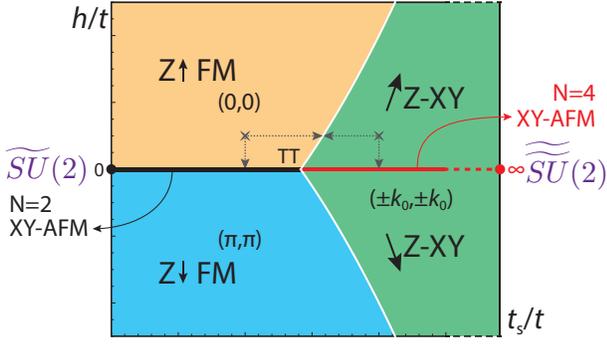}
    \caption{ The global phase and phase transitions as a function of $h/t$ and $ t_s/t $.
    In the yellow region, the mean field state is the Z $ \uparrow $ FM-SF with only one minimum located at $(0,0)$.
    In the blue region, the mean field state is the Z $ \downarrow $ FM-SF with only one minimum located at $(\pi,\pi)$.
    Both states keep the $ [C_4 \times C_4]_D $ symmetry, so are non-degenerate.
    In the green regime, the mean field state is the Z-XY FM-SF with four minima located at $(\pm k_0,\pm k_0)$
    where $ k_0 $ depends on $ t, t_s, h $. It breaks $ [C_4 \times C_4]_D \to 1 $, so is 4 fold degenerate.
    There is a first order quantum phase transition at $ h=0 $ driven by the roton dropping tuned by
    the Zeeman field.  There is a second order quantum bosonic Lifshitz phase transition with the dynamic exponents
    $ z_x=z_y=2 $  driven by the softening of the SF Goldstone mode tuned by the ratio $ t_s/t $ across the phase boundary.
    The two phase transition boundaries meet at the Topological Tri-critical (TT) point at $ h=0, t_s/t=\sqrt{2} $ which separates
    the $ N=2 $ XY-AFM state from  the $ N=4 $ XY-AFM state.
    Their boson condensation functions are given by  Eq.\ref{deg} and Eq.\ref{k1k30},\ref{k2k40} respectively.
    Both states have the same XY-AFM spin-orbital structure, respect the anti-unitary $ Z_2 $ symmetry and
    break the $ [C_4 \times C_4]_D $ symmetry, so they can be distinguished only by
    the different topology of the BEC condensation momenta in Fig.\ref{tpt} instead of by any differences in the symmetry breaking patterns.
    The ground state at the left Abelian point is a FM SF in the $ \tilde{\mathbf{S}} $ basis,
    That at the right Abelian point is a frustrated SF with the coplanar $ 90^{\circ} $ spin-orbital structure  in the $ \tilde{\tilde{\mathbf{S}}} $ basis.
    For the finite temperature phase transitions above all the SF phases, see Fig.\ref{rotondrive},\ref{sfdrive}.}
\label{phasedia}
\end{figure}

However, so far the experiment\cite{2dsocbec}  is still at single particle level: namely mapping of the topological
 band of the QAH model Eq.\ref{qah}
using the thermally excited  $ ^{87}$Rb  atoms, no study of quantum many body phenomena yet.
The ultimate goal of the experiments in \cite{2dsocbec} is to study many body phenomena of SOC $ ^{87}$Rb  BEC.
This directly motivated us to investigate possible many-body phenomena of the QAH model of the spinor bosons  Eq.\ref{qah}
at a weak interaction.
We find the competition among the hopping, the SOC and the Zeeman field in the QAH model Eq.1 leads to
a hopping dominated ( Abelian )  regime and a SOC dominated ( non-Abelian ) regime in the phase diagram of Fig.1.
While the upper part with the Zeeman field $ h > 0 $ is related to the lower part $ h < 0 $ by a anti-unitary $ Z_2 $ Reflection transformation Eq.\ref{R}. There are
 various novel spin-bond correlated superfluid states and also new classes of quantum or topological  phase transitions among these SF phases.
 One transition is a first order one driven by roton droppings ( but with non-vanishing roton gaps $ \Delta_R $ )
 at $ h=0 $ tuned by the Zeeman field $ h $.
 Another is a second order one driven by the softening of the superfluid Goldstone mode tuned by the ratio of
 SOC strength over the hopping strength. It is a bosonic Lifshitz transition with the dynamic exponents $ z_x=z_y=2 $
 and an accompanying $ [C_4 \times C_4]_D $ symmetry breaking.
 The two phase boundaries meet at the topological tri-critical ( TT )  point which separates the $ h=0 $ line into two SF phases
 with $ N=2 $ and $ N=4 $ condensation momenta respectively.
 At  $ h =0 $, the system has the  $ Z_2 $ Reflection symmetry,
 there are classically degenerate family of states with the dimension $ 3 $ and $ 5 $ manifold in the two sides respectively.
 We perform a novel systematic order from quantum disorder analysis to find true quantum ground states
 and also calculate the roton gaps $ \Delta_R $ generated by the order from disorder mechanism on both sides of the TT point.
 The two SF phases have the same spin-orbital XY-AFM spin-orbital structure with $ d=4 $ degeneracy,
 respect the anti-unitary $ Z_2 $ symmetry
 and break the identical other symmetries of the Hamiltonian. So they can not be distinguished
 by any difference in the symmetry breakings
 except by the different topology of the BEC condensation momenta.
 This could be a first bosonic analog of the fermionic topological Lifshitz  transition
 of free fermions with Fermi surfaces, Dirac points or Weyl points \cite{topo1,topo2}.
 However, when moving away from $ h=0 $ line, the $ Z_2 $ Reflection  symmetry is lost, the TT is converted to
 the  second order bosonic Lifshitz transition with the accompanying $ [C_4 \times C_4]_D $ symmetry breaking.

\begin{figure}[!htb]
\centering
    \includegraphics[width=0.35 \textwidth]{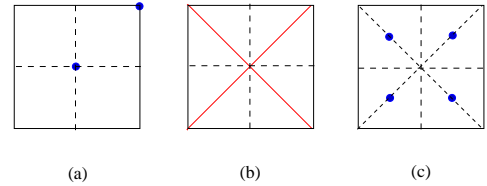}
    \caption{ The $ Z_2 $ reflection symmetry protected TPT at $ h=0 $. (a) The SF at the left has $ N=2 $ BEC condensation momenta
    at $ (0,0) $ and $ ( \pi,\pi) $.
    (b) At the TPT, the dispersion becomes flat along the two crossing lines $ k_x=\pm k_y $.
    (c) The SF at the right has $ N=4 $ BEC condensation momenta at $ ( \pm \pi/2, \pm \pi/2) $.
    Both SF phase have the same XY-AFM spin-orbital structure with
    $ d=4 $ degeneracy \cite{naive}.}
\label{tpt}
\end{figure}

The results achieved in this work can be detected in the ongoing experiment \cite{2dsocbec,more},
especially the topological phase transition at the TT can be directly detected by the Time of flight (TOF) imaging.
They may guide and inspire the ongoing world wide experimental efforts \cite{expk40,expk40zeeman,SD,clock,clock1,clock2} to
study new many body phenomena due to the interplay between SOC and interactions.
Surprisingly, so far, there is very little work to study the effects of interaction on QAH in the materials side either,
so our results and methods may also shed some lights on studying interaction effects on QAH materials \cite{QAHthe,QAHexp}.

  The experimentally realized Quantum Anomalous Hall ( QAH ) model of spinor bosons in a square lattice
  is described by the Hamiltonian:
\begin{eqnarray}
   H_{QAH} & = &	-t\sum_{\langle ij\rangle_{x,y} }
	(a^\dagger_{i\uparrow}a_{j\uparrow}
	-a^\dagger_{i\downarrow}a_{j\downarrow})
	-h \sum_i (n_{i\uparrow}-n_{i\downarrow})    \nonumber    \\
	& + & \sum_i[ it_{s}(a^\dagger_{i} \sigma_x a_{i+x}
	+ a^\dagger_{i} \sigma_y a_{i+y})+h.c.]       \nonumber    \\
   & + & \frac{U}{2} \sum_{i} ( n^2_{i \uparrow}+  n^2_{i \downarrow} + 2 \lambda  n_{i \uparrow} n_{i \downarrow} )
 - \mu \sum_{i} n_i
\label{qah}
\end{eqnarray}
  where the spinor $ a_{i \alpha} $ could be either fermions or bosons.
  For $ ^{87}$Rb  atoms, the two pseudo-spin components $ \alpha=\uparrow, \downarrow $
  denote the two hyperfine states $|1,m_F=0\rangle$ or $|1,m_F=-1\rangle$.
  In this paper, we focus on the spin isotropic interaction $ \lambda=1 $ which may also be the
  most experimentally relevant  one.

  In addition to the particle number $ U(1)_c $ symmetry,
  the Hamiltonian has a $ [ C_4 \times C_4]_D $ symmetry. At $ h=0 $, there is an enlarged anti-unitary $ Z_2 $ Reflection symmetry:
\begin{equation}
   R= (-1)^{i} R_z(\pi) \mathcal{T}
\label{R}
\end{equation}
 where (1) A spin rotation $R_z(\pi)$ leads to $t_{s}\to-t_{s}$;
 (2) A time reversal $\mathcal{T}$ leads to $(t, h )\to(-t,-h )$;
 (3) A sublattice rotation $ c_{i \alpha} \to (-1)^{i} c_{i \alpha} $ leads to $(t,t_{s})\to(-t,-t_{s})$.
 Under $ R $, $ h \to -h $.
 So there is an enlarged $ R $ symmetry at $ h=0 $.





{\bf RESULTS }


{\bf 1. Mean field Analysis of the ground states in $ h, t_s/t $ space.  }

  In the weak interaction limit $ U \ll t $ which is also the most  experimentally easily accessible limit,
  One can first diagonals the non-interacting $ U=0 $ the bosonic QAH Hamiltonian Eq.\ref{qah} and locate the minima positions
  shown in Fig.\ref{phasedia} which has 3 different regimes shown in Fig.1.

   (a) When $h/t>0$ and  $h/t>2t_{\rm s}^2/t^2-4$,  there is only one minimum located at $(0,0)$ with the
   spinor spin up, called Z $ \uparrow $ FM-SF. It respects the $ [C_4 \times C_4]_D $ symmetry.

   (b) When $h/t<0$ and  $h/t<4-2t_{\rm s}^2/t^2$,  there is only one minimum located at $(\pi,\pi)$
      with the spinor spin down, called Z $ \downarrow $ FM-SF. It also respects the $ [C_4 \times C_4]_D $ symmetry.

   (c) When $4-2t_{\rm so}^2/t^2<h/t<2t_{\rm so}^2/t^2-4$,
there are four minima:
$\mathbf{K}_1=(k_0,k_0)$, $\mathbf{K}_2=(-k_0,k_0)$,
$\mathbf{K}_3=(-k_0,-k_0)$, and $\mathbf{K}_4=(k_0,-k_0)$,
where $k_0=\arccos\left(\frac{th}{2t_s^2-4t^2}\right)$;
the corresponding spinors are
\begin{align}
    \chi_{n}=\begin{pmatrix}
		-e^{-i\phi_n}\sin(\theta_n/2)\\
		\cos(\theta_n/2)
	     \end{pmatrix}
\label{spinorn}
\end{align}
where  $  n=1,2,3,4 $ and
\begin{align}
    \theta_n  & =  \arccos\left[
    \frac{-h-4t\cos k_0}
	 {\sqrt{(h+4t\cos k_0)^2+8t_s^2\sin^2 k_0}}\right],\quad  \nonumber  \\
    \phi_n  & =  \left(n-\frac{1}{2}\right)\frac{\pi}{2},
\label{twonn}
\end{align}

 In the limit $ t \to 0 $ ( namely, close to the right Abelian axis ), the phase boundary $ h \sim  2 t^{2}_s/t $ diverges,
 it means there is no phase transition on the right axis in Fig.1. Indeed, as $ t \to 0 $, $ k_0 \to \pi/2 $, $ \cos \theta_n= \frac{h}{ h^2 + 8 t^2_s } < 1 $, so $ \theta_n \to 0 $ only in the limit $ h \to \infty $, indicating no phase transition on the right axis.
 This is in sharp contrast to the strong coupling limit  where it was established
 there is a quantum phase transition at a finite Zeeman field \cite{rhht}.

  The most general single-particle ground state can be written as:
\begin{align}
    \Psi= \sum_{i=1}^4  c_i e^{iK_i r_i}\chi_i
\label{fourPW}
\end{align}
 where the four coefficients are normalized $\sum_{i=1}^4|c_i|^2=1$.
 Its interaction  energy can be written in terms of $ c_i $. When $\lambda=1$,
 its minimization shows the mean field ground state is always a plane wave state where the interaction energy simplifies to
 $   E^{0}_\text{int}=N_sU  $. The spin of the plane wave state at $ \mathbf{K}_n $  is along  $ ( \theta_n, \phi_n) $
 which has both Z component and a component in XY plane oriented at $ \phi_n=(n-\frac{1}{2} )\pi/4 $, so named
 Z-XY FM-SF. It breaks the $ [C_4 \times C_4]_D $ symmetry.

 As $ h \to 0 $, then $ k_0 \to \pi/2 $,  $ \theta_n \to \pi/2 $, so the spin lies in the XY plane.
 As $ h $ approaches the upper ( or lower ) boundary, $ k_0 \to 0 $ ( or $ k_0 \to \pi $ ),
 $ \theta_n \to 0 $, so the spin aligns along the spin up  ( or spin down ) direction, so the Z-XY FM-SF reduces to the
 Z $ \uparrow $ FM-SF ( or  Z $ \downarrow $ FM-SF ) respectively.

 At the Tri-critical point T at $ (h=0, t_\text{s}/t=\sqrt{2} ) $, the minima become
 the two crossing lines $ k_x \pm k_y=0 $.

  When $ h \neq 0 $, these mean field states are labeled as Z $ \uparrow $ FM-SF,  Z $ \downarrow $ FM-SF and Z-XY FM-SF
  in the phase diagram Fig.1.  In the following sections, we will treat $ U $ as a weak interaction to study the excitations above all the phases.
  However, at $ h=0 $, there are classically degenerate family of states,
  so mean field can not determine the true quantum ground state.
  In Sec.3 and 5, we will perform the Order from quantum disorder analysis to determine the true ground state
  from the degenerate family of classical states at $ h=0 $, then calculate the excitation spectrum.
  We will also investigate quantum or topological phase transitions among all these phases.

{\bf 2. The excitation spectrum in the  Z $ \uparrow $ FM-SF  at $ h \neq 0, |h/t|>2t_s^2/t^2-4$. }

  When the spectrum minimum is located at $(0,0)$ or $(\pi,\pi)$, the non-interacting Hamiltonian takes a simple form
\begin{align}
    H_0(k)=-[h+2t(\cos k_x+\cos k_y)]\sigma_z
\end{align}
   where the SOC $ t_s $ drops out.

   Let us consider $h>0$ case, thus the minimum is located at $(0,0)$.
In the weak coupling limit $ U/t \ll 1 $, by writing
\begin{align}
	a_{k\uparrow}\to\sqrt{N_0}\delta_{k,0}+\psi_{k\uparrow},\quad
	a_{k\downarrow}\to\psi_{k\downarrow}
\end{align}
  where $N_0$ is the number of condensate atoms,
  and $N_s$ is the total number of lattice sites, thus $n_0=N_0/N_s$ is the condensate fraction.

   We can perform the expansion $  \mathcal{H}=\mathcal{H}^{(0)}+\mathcal{H}^{(1)}+\mathcal{H}^{(2)}+\cdots
    $ where the superscript denotes the order in the quantum fluctuations.
    The zeroth order term $ \mathcal{H}^{(0)}=E_0= -\frac{1}{2}U n_0N_0 $  is the classical energy of the condensate.
  The vanishing of the linear term sets the value of the chemical potential $\mu=-h-4t+U n_0$.
  Diagonizing  $ \mathcal{H}^{(2)} $ by a generalized $ 4 \times 4 $ Bogliubov transformation leads to:
\begin{align}
     H=E_{0}+ E^{(2)}_0 + \sum_{k,s=\pm} \omega_s(k)( \alpha_{sk}^\dagger\alpha_{sk} + 1/2)
\label{e0e2modes}
\end{align}
where  $ E^{(2)}_0=-( h+4t+ \frac{1}{2} U n_0 ) $. $ \omega_{\pm}(k) $ are the two Bogliubov modes.

\begin{figure}[!htb]
\centering
\includegraphics[width=\linewidth]{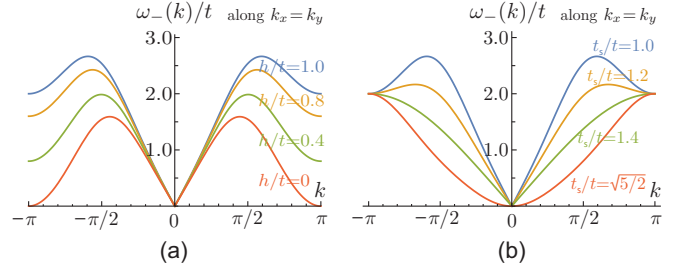}
\caption{
The Critical Behaviour of the superfluid Goldstone mode $\omega_-(k_x=k_y)$ starting at $ ( t_s/t=1, h/t=1 ) $.
(a) Decreasing $h/t$ from 1 to 0, the roton at $ \vec{Q}= (\pi, \pi) $ gets lower and lower and drops to zero at $ h=0 $
signifying a first order transition.
(b) Increasing $t_\text{s}/t $ from $1$ to its critical value $\sqrt{5/2}$, the SF Goldstone mode gets softer and softer
and becomes quadratic  at the phase boundary signifying a second order bosonic transition.
We used $n_0U/t=1$.}
\label{roton1}
\end{figure}

Since $\omega_+(\mathbf{k})>\omega_-(\mathbf{k})$ always holds, so we focus on $\omega_-(\mathbf{k})$
which displays a gapless superfluid Goldstone mode near $ \mathbf{k}=(0,0) $ and
a gapped roton mode at $ \mathbf{k}=(\pi,\pi) $. In the long wavelength limit, we find
the superfluid Goldstone mode becomes linear $
    \omega_-(k)=c|\mathbf{k}| $ where the velocity is
\begin{align}
    c=\sqrt{\frac{2(4t^2-2t_\text{s}^2+th)Un_0}{4t+h}}
\label{slope}
\end{align}
  which is shown in Fig.\ref{roton1}a. In $ h \to 0 $ limit, it reduces to $ c= \sqrt{ n_0 U t (2- t^{2}_s/t^2) } $.
  How the SF Goldstone mode and the gapped roton mode behave as one approaches
  the phase boundaries in Fig.1 is shown in Fig.\ref{roton1}.

{\sl (a). The second order bosonic Lifshitz transition from the Z FM to the Z-XY FM }

  For a general $ h $,  it is easy to see that the velocity of the SF Goldstone mode Eq.\ref{slope}
  vanishes as one approaches the phase boundary ( Fig.\ref{roton1}a ) where the dispersion becomes quadratic:
\begin{align}
	\omega_-(\mathbf{k}) =\sqrt{n_0Ut\left[\frac{1}{2}(k_x^4+k_y^4)-\frac{t^2}{2t_s^2}(k_x^2+k_y^2)^2\right]}
\label{lif}
\end{align}
   which shows the transition from the Z FM to the Z-XY FM is a bosonic Lifshitz transition with the dynamic
   exponents $ z_x=z_y=2 $. The equality of $ z_x=z_y $ is dictated by the $ [C_4 \times C_4]_D $ symmetry in the Z FM SF
   with either spin up or spin down in Fig.1 \cite{z2}.
   There is an accompanying transition in the spin sector from the Z FM to the Z-XY FM with the symmetry breaking
   pattern $ [C_4 \times C_4]_D \to 1 $.
   The Bosonic Lifshitz transition from commensurate canted phase to non-coplanar IC-SkX with the anisotropic dynamic exponents $ (z_x=1, z_y=3 ) $ was studied in \cite{rhh}.

  In Fig.\ref{roton1}a, at a fixed $ h=1 $, as $ t_s $ increases to $ t_s=\sqrt{5/2} $ sitting on the
  boundary between $ (0,0) $  Z $ \uparrow $ FM-SF and
  the  Z-XY FM-SF at $ (\pm k_0, \pm k_0) $, the slope of the Goldstone mode  at $ (0,0) $ decreases as in Eq.\ref{slope},
  then becomes quadratic at the boundary as shown in Eq.\ref{lif}.

{\sl (b). The first order transition from the Z $ \uparrow $ FM-SF to the Z $ \downarrow $ FM-SF driven by the roton dropping. }

  In Fig.\ref{roton1}b, at a fixed $ t_s=1 $, as $ h $ increases to $ 0 $ sitting on the boundary between  Z $ \uparrow $ FM-SF at $ (0,0) $ and
  the Z $ \downarrow $ FM-SF at $ (\pi,\pi) $, the roton mode gets lower and lower,
  then touches zero at $ \vec{Q}= (\pi,\pi) $.
  In the small $ h $ limit, we find the roton gap $\Delta_R= 2 |h| $ which is independent of $ t_s $
  ( the dashed line in Fig.\ref{roton1gap}a ).
  Naively, this behaviour may signify a possible second order transition.
  However, as to be shown in the next section, the order from quantum disorder analysis show that
  the quantum fluctuations will open a roton gap at $ h=0 $ ( the real line in Fig.\ref{roton1gap}a ).
  So the roton driven transition is a first order one.

{\sl (c). The TT point approaching from the left. }

   The above two phase transition boundaries meet at the Topological Tri-critical (TT)
   point $ ( h=0, t_s=\sqrt{2}t ) $ where
\begin{align}
   \omega_-(k)=\sqrt{tn_0U}/2 | k_x^2-k_y^2 |
\label{cross}
\end{align}
which becomes flat along the two crossing lines $ k_x = \pm k_y $ shown in Fig.\ref{tpt}b.


{\bf 3. The Order from disorder phenomena and the $ N=2 $ XY-AFM state at $ h=0, t_s/t < \sqrt{2} $. }

  At the two minima $ (0,0) $ and $ ( \pi,\pi) $, the $ t_s $ term vanishes, so the effects of SOC do not show up at the mean field level.
  They only show up at quantum fluctuations. At $ h > 0 $ ( or $ h < 0 $ ), the minimum is at $ (0,0) $  ( or  $ ( \pi,\pi) $ ),
  it is these quantum fluctuations which lead to the gapless Goldstone mode
  and the gapped roton mode discussed in the above section.

  At $ h=0 $, in the left of the T point $ 0 < t_s/t < \sqrt{2} $ in Fig.1,
  the two minima become degenerate, there is a classically degenerate family of ground states:
 \begin{equation}
  \Phi_{0,L}
	= \sqrt{N_0} [ c_0
	\begin{pmatrix}
		1\\
		0
	\end{pmatrix} + c_{\pi} (-1)^{x+y}
	\begin{pmatrix}
		0\\
		1
	\end{pmatrix}]
\label{deg}
\end{equation}
 where $ c_0 $ and $ c_{\pi} $ are any two complex numbers satisfying the normalization condition
 $ | c_0 |^2 + |c_{\pi}|^2=1 $.  The "quantum order from disorder" mechanism is needed to determine that the quantum ground state
 in this family of the classical degenerate ground state.


We write the spinor field as the  condensation part Eq.\ref{deg} plus a quantum fluctuating part
$ \Psi= \sqrt{N_0} \Psi_0+ \psi $. Again, the zero order term gives the classical ground state
energy $ E_0=-\frac{1}{2}Un_0N_0 $.
Setting the linear term vanish gives the value of the chemical potential $\mu=-4t+Un_0$.
Diagonizing  $ \mathcal{H}^{(2)} $ by a generalized $ 8 \times 8 $  Bogliubov transformation leads to:
\begin{align}
    H^{(2)}
	= E_\text{GS}[c_0,c_\pi]  +\hspace{-0.2cm}\sum_{n,k\in\text{RBZ}}\omega_n(k)\alpha_{n,k}^\dagger\alpha_{n,k}
\label{ofdleft}
\end{align}
where the RBZ has the diamond shape $ | q_x+ q_y | \leq \pi $ and
$\omega_n(k)$ with $n=1,2,3,4$ are the four Bogoliubov modes and the
 ground-state energy incorporating the quantum fluctuations is
\begin{align}
    E_\text{GS}[c_0,c_\pi]= E_{0t}+\frac{1}{2} \sum_{n,k\in\text{RBZ}}\omega_n(k)
\end{align}
    where $ E_{0t}= E_0 - ( 4t + \frac{1}{2}Un_0 ) $.

  After parameterizing $c_0$ and $c_\pi$ as
\begin{align}
    c_0=\cos(\theta/2),\quad c_\pi=e^{i\phi}\sin(\theta/2).
\end{align}

 We find the minima of $ E_\text{GS}[c_0,c_\pi] $ is located at
 $ \theta=\pi/2 $ and $ \phi=\pi/4, 3\pi/4, 5 \pi/4, 7 \pi/4 $ (See also Fig.\ref{ABfigure} ).  So the ``order from disorder'' mechanism
 picks up the 4-fold degenerate state in Eq.\ref{deg} as the quantum ground states whose associated
 spin-orbit structure is:
\begin{align}
    \mathbf{S}_i=\langle \frac{\hbar}{2}\vec{\sigma}\rangle
    =\frac{\hbar}{2}\big( \pm (-1)^{x+y}/\sqrt{2},\pm (-1)^{x+y}/\sqrt{2},0\big)
\label{AFMXY1}
\end{align}
which is a AFM state in the XY plane. Obviously, it breaks the $ [C_4 \times C_4]_D \to 1 $ with $ d=4 $ fold degeneracy.

\begin{figure}[!htb]
\centering
\includegraphics[width=0.48\linewidth]{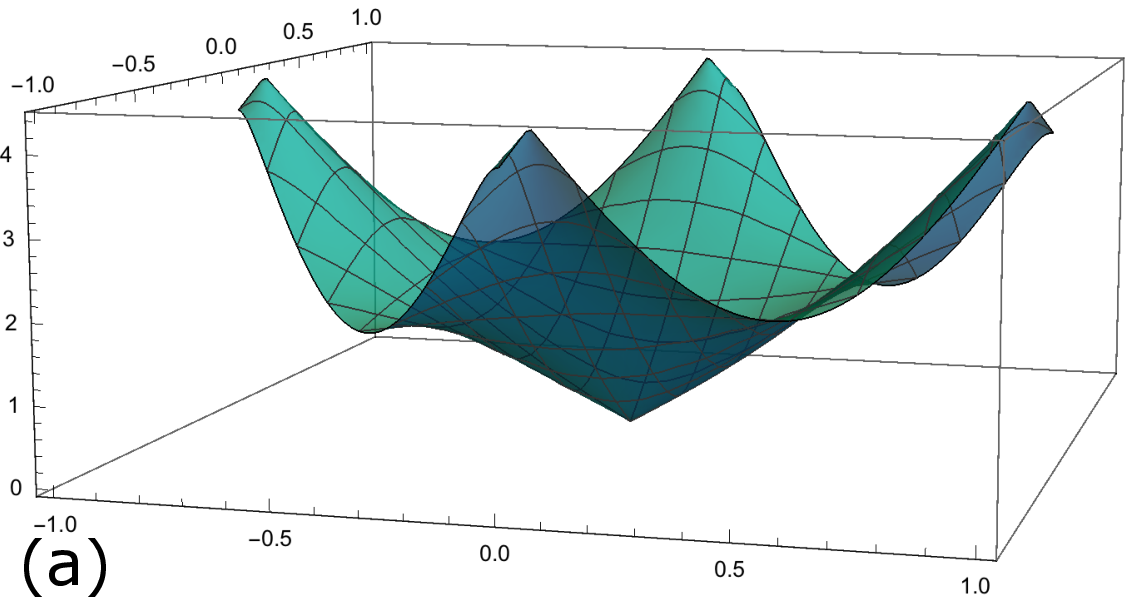}
\enskip
\includegraphics[width=0.48\linewidth]{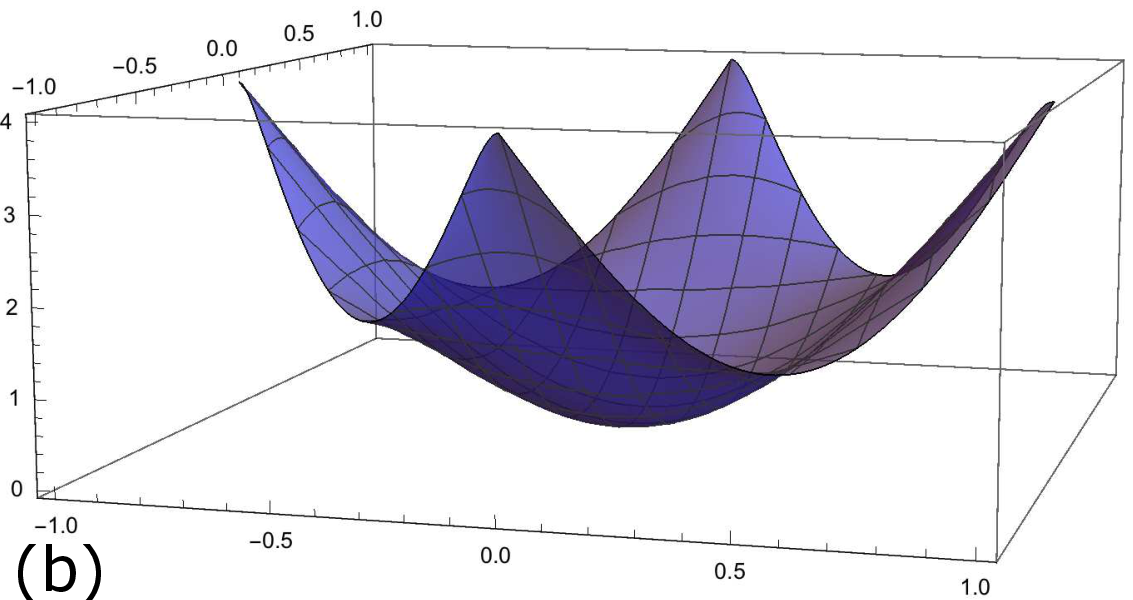}
    \caption{(a) There is a linear SF Goldstone mode $\omega_1(k)$ at $(0,0)$.
	(b) There is also a quadratic roton mode $\omega_2(k)$ at $(0,0)$.
     The roton mode will acquire a gap due to the order from disorder mechanism as shown in Fig.\ref{roton1gap}.
	$\omega_3(k)$ and $\omega_4(k)$ are fully gapped higher energy modes, so not shown.
	We used $n_0U/t=1$ and $t_s/t=1/2$.}
\label{GR}
\end{figure}

\begin{figure}[!htb]
\centering
    \includegraphics[width=\linewidth]{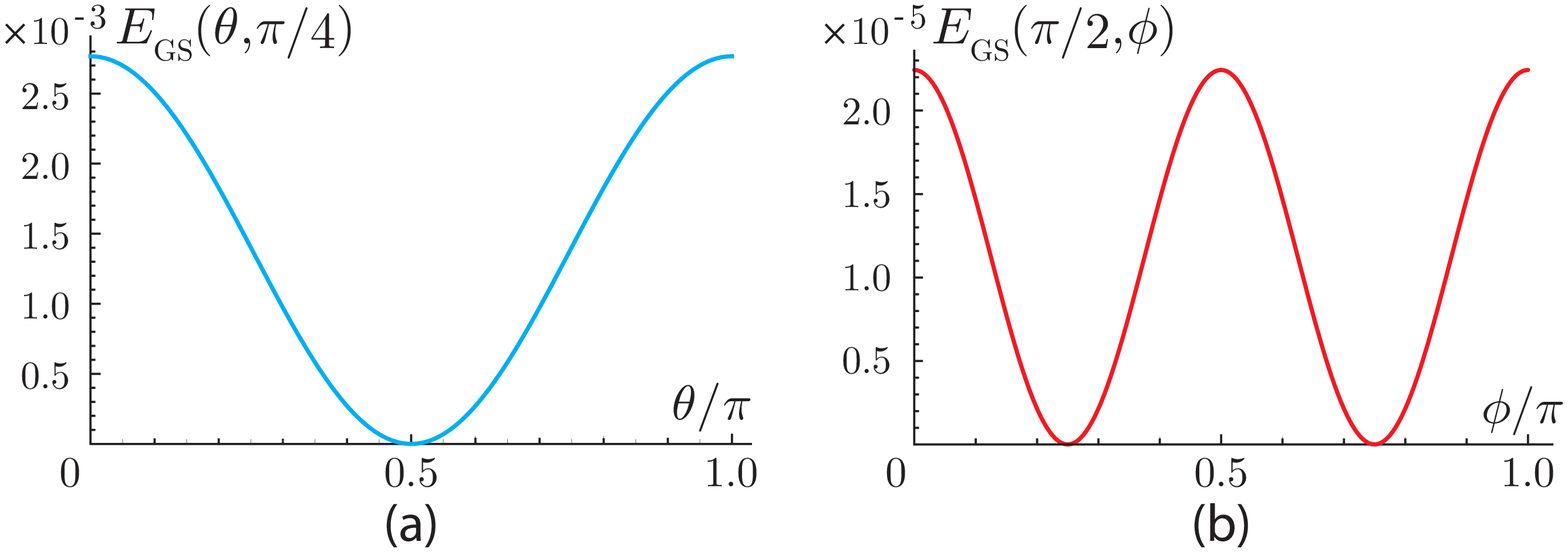}
    \caption{The quantum ground-state energy near its minimum at
    (a)  $\theta=\pi/2 $ at fixed $ \phi=\pi/4 $, (b) $ \phi=\pi/4, 3\pi/4 $ at fixed  $\theta=\pi/2 $
    where the coefficients $ A $ and $ B $ in Eq.\ref{AB} can be extracted respectively.
    We used $n_0U/t=1$ and $t_s/t=1$. }
\label{ABfigure}
\end{figure}

\begin{figure}[!htb]
\centering
\includegraphics[width=\linewidth]{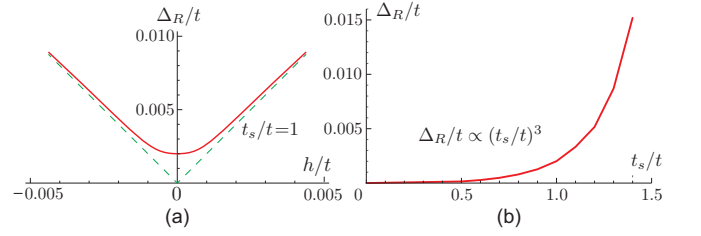}
	\caption{(a)Roton gap computed
	by the order from disorder analysis as a function of zeeman field $h$ at a given $ t_s/t >1 $.
	Other parameters are $n_0U=1$, $n_0\approx1$.
	The dashed line of the roton dropping is before the order from disorder calculations.
	(b) At $ h=0 $, the Roton gap computed by the order from disorder analysis
        is an increasing function of  $t_\text{s}$ as approaching to the T point from the left.
	Other parameters are $n_0U=1$, $n_0\approx1$.
        At $ t_s=0 $, it is nothing but the FM spin wave mode.
        At small $ t_s $, it can also be calculated by the perturbation theory in $ t_s $
        which leads to $ \Delta_R \sim t^{3}_s $ ( see Method ).
        Compare with $ \Delta_{3R} $ on the right side in Fig.\ref{roton2gap}. }
\label{roton1gap}
\end{figure}

After identifying the correct quantum ground-state as the $ N=2 $ XY-AFM state,
we can also evaluate all $\omega_{1,2,3,4}(k)$ in Eq.\ref{ofdleft}.
There are one linear $\omega_1 $ SF Goldstone mode and
one $\omega_2 $ quadratic roton mode located at $(0,0)$ shown in Fig.\ref{GR}.


{\sl (a) The Roton mass gap generated from the "order by disorder" mechanism in the $ N=2 $ XY-AFM state }

The "order from disorder" mechanism picks up $\theta=\pi/2$ and $\phi=\pi/4$ as the quantum ground state.
Then we can expand the ground-state energy around the minimum as:
\begin{align}
	E_\text{GS}[\theta,\phi]=E_0+\frac{A}{2}\delta\theta^2+\frac{B}{2}\delta\phi^2
\label{AB}
\end{align}
where $\theta=\pi/2+\delta\theta$ and $\phi=\pi/4+\delta\phi$, the two coefficients $ A \sim (n_0 U)^2/t $ and
$ B \sim (n_0 U)^2/t $ can be extracted from Fig.\ref{ABfigure}.

  Using the commutation relations $ [\frac{1}{2}n_0\delta \theta, \phi ]= i \hbar $, one can see that
  the "quantum order from dis-order " mechanism generates the roton gap:
\begin{equation}
  \Delta_R=2\sqrt{AB}/n_0
\label{deltar}
\end{equation}
  which is shown in the Fig.\ref{roton1gap}.
  The SF Goldstone mode remains unaffected. This is consistent with its protection by the $ U(1) $ symmetry breaking.

  In fact, the $ R $ symmetry at $ h=0 $ dictates $ |c_0|^2=|c_{\pi}|^2=1/2 $ without fixing the relative phase.
  So the $ N=2 $ XY-AFM states still keep the $ R $ symmetry.
  The transition at $ h=0 $ is a first order phase transition driven by the roton dropping tuned by the Zeeman field $ h $.
  But the roton still has a non-zero gap $ \Delta_R $  before it sparks the first order transition as shown in Fig.\ref{roton1gap}.
  The $ N=2 $ XY-AFM state is a pure state instead of a mixed state of the $ (0,0) $ spin up state and $ (\pi,\pi) $ spin down state
  with any ratio. This is in sharp contrast to the 1st order transition between the Y-x state and X-y state in Ref.\cite{rhrashba}.
  However, as to be shown in Fig.\ref{rotondrive}, it may melt into such a mixed state through a finite temperature
  $ Z_4 $ clock transition.

{\bf 4. The excitation spectrum in the Z-XY FM SF at $ h \neq 0, |h/t|<2t_s^2/t^2-4 $. }

AS shown in Sec.1, the ground state is just a plane wave (PW) state.
Without a loss of generality, we choose the PW state at the momentum $ \mathbf{K}_1$.
From Eq.\ref{spinorn}, we can rewrite spinor field as:
\begin{align}
    \begin{pmatrix}
	b_{k\uparrow}\\
	b_{k\downarrow}\\
    \end{pmatrix}
    =
    \sqrt{N_0}
    \begin{pmatrix}
	-e^{-i\phi}\sin(\theta/2)\\
	\cos(\theta/2)\\
    \end{pmatrix}
    \delta_{k, \mathbf{K}_1 }
    +
    \begin{pmatrix}
	\psi_{k\uparrow}\\
	\psi_{k\downarrow}\\
    \end{pmatrix}
\label{zxy}
\end{align}

Following the similar procedures as in section 2,
Setting the Linear term vanishing determines the chemical potential $\mu=-\omega_0+Un_0$ where
$  \omega_0=\omega_-^\text{free}( \mathbf{K}_1)=\sqrt{(h+4t\cos k_0)^2+8t_\text{s}^2\sin^2 k_0}  $.
We also reach the form of Eq.\ref{e0e2modes} where
$ E_0= -\frac{1}{2}Un_0N_0,
  E_2=\frac{1}{2}\sum_k[\omega_+(\mathbf{k})+\omega_-(\mathbf{k})-2\omega_0-Un_0] $.
 We found  one Goldstone mode at $\mathbf{K}_1=(k_0,k_0)$,
 which, in the long wavelength limit, takes the form
\begin{align}
    \omega_-(\mathbf{K}_1\!+\!\mathbf{q})
    \sim\sqrt{v_{\parallel}^2(q_x+q_y)^2+v_{\bot}^2(q_x-q_y)^2},
\end{align}
where $v_{\parallel}$ and $v_{\bot}$ take the analytic form:
\begin{align}
    v_{\parallel}^2
    &=
	\frac{t_\text{s}n_0U}{\sqrt{8-\frac{h^2}{2t^2-t_s^2}}}
	\left[2-\frac{t^2}{2t_s^2}\left(8-\frac{h^2}{t_s^2-2t^2}\right)\right]\\
    v_{\bot}^2
    &=\frac{t_\text{s}n_0U}{\sqrt{8-\frac{h^2}{2t^2-t_s^2}}}
	\left[2-\frac{1}{2}\left(\frac{ht}{t_s^2-2t^2}\right)^2\right]
\end{align}

{\sl (a). The second order bosonic Lifshitz transition from the Z-XY FM to the Z FM }

   It is easy to see that both velocities vanish in the phase boundary in Fig.1 where the dispersion
   becomes quadratic which is identical to that listed in Eq.\ref{lif}.
   This agreement between approaching the boundary from the left and
   from the right shows that  the transition is indeed a second order bosonic Lifshitz
   transition with the dynamic exponent $ z_x=z_y=2 $ and also the accompanying symmetry breaking $ [C_4 \times C_4]_D  \rightarrow 1 $ in the spin sector.

 {\sl (b). The first order transition between the two Z-XY FM states driven by the roton dropping. }

   In the  $ h \to 0$ limit, $ k_0 \to \pi/2 $, the Goldstone mode becomes
\begin{align}
    \omega_{1}=\sqrt{\frac{\sqrt{2}n_0U}{t_\text{s}}
	    [t_\text{s}^2(q_x^2+q_y^2)-t^2(q_x+q_y)^2]}
\label{gold1}
\end{align}
  where $ \vec{k}= \mathbf{K}_1+\mathbf{q} $.

\begin{figure}[!htb]
\centering
\includegraphics[width=\linewidth]{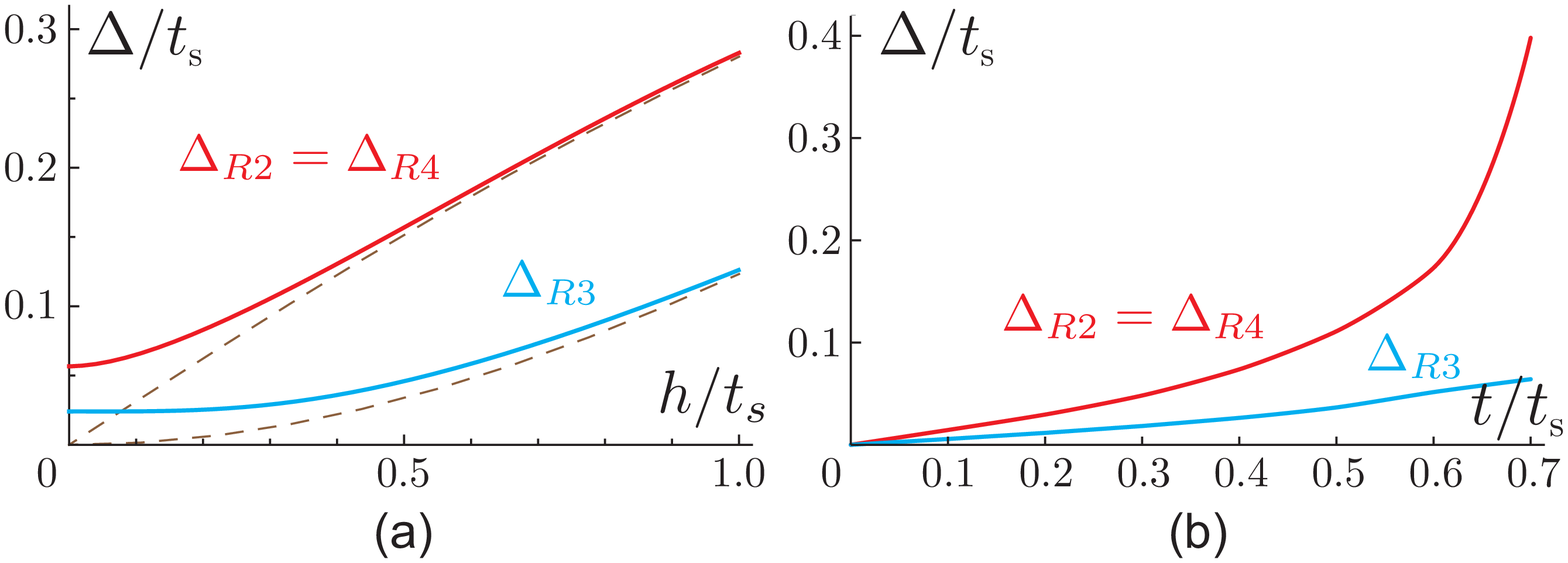}
    \caption{ (a) The three roton gaps $\Delta_{2R}=\Delta_{4R} \gg \Delta_{3R} $
	computed from the order from disorder analysis
     as a function of $h/t_s$  at a fixed $t/t_s =1/3 $. We used $n_0U/t_s=1$.
     The dashed line is the roton dropping without the order from disorder analysis.
	(b) All the three roton gaps are monotonically increasing functions of $t$ as approaching to
    the TT point from the right at $h=0$.  We used $n_0U=t_s$.
    Compare $ \Delta_{3R} $ with $ \Delta_{R} $ on the left side in Fig.\ref{roton1gap}.
    The first order transition at $ t_s/t > \sqrt{2}, h=0 $ is driven by the roton dropping at
    $ (\pi,\pi) $ with the gap $ \Delta_{3R} $.
    $ \Delta_{2R} = \Delta_{4R}  $ are essentially irrelevant to the first order transition. }
\label{roton2gap}
\end{figure}
\begin{figure}[!htb]
\centering
    \includegraphics[width=0.22\textwidth]{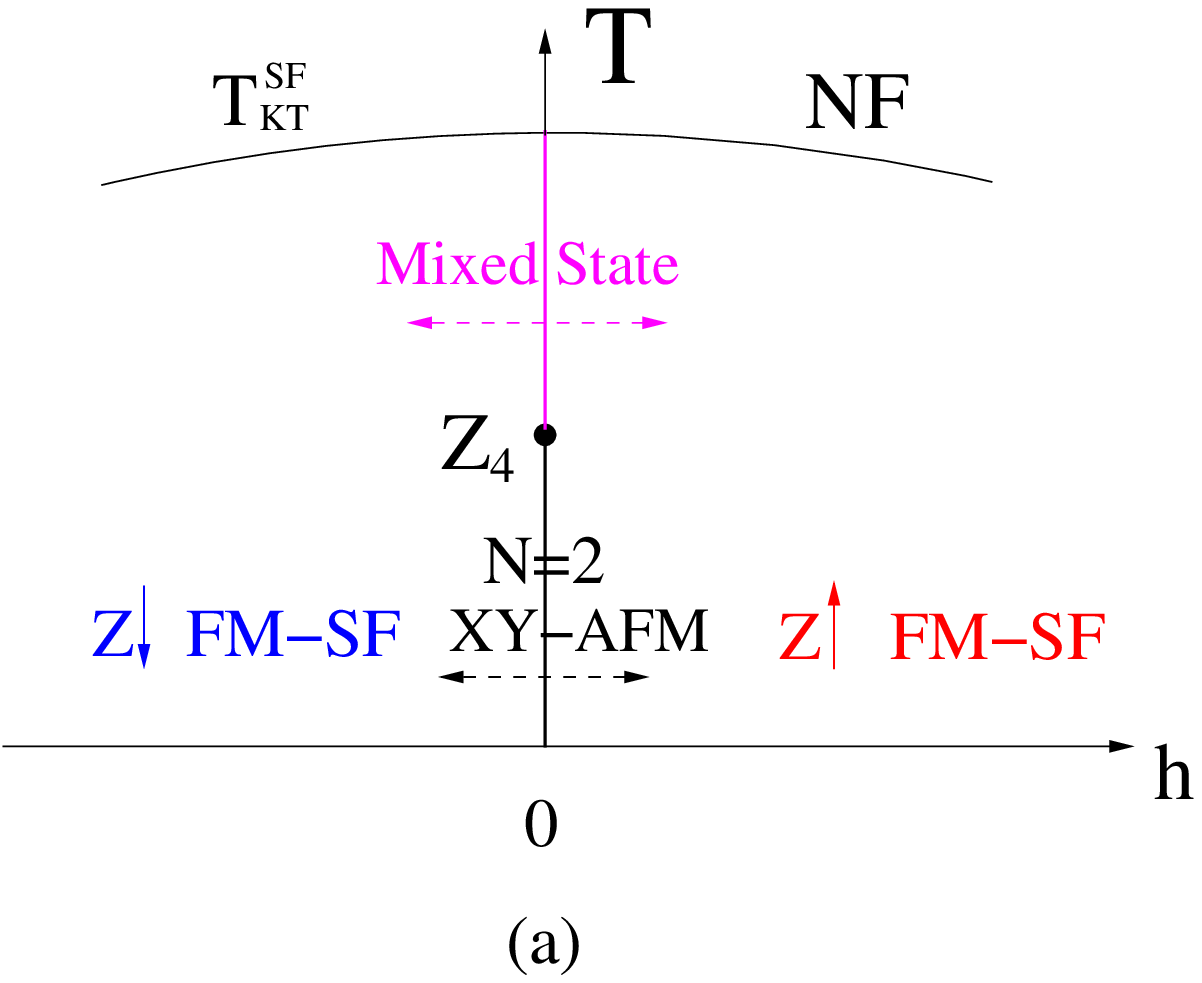}
    \includegraphics[width=0.22\textwidth]{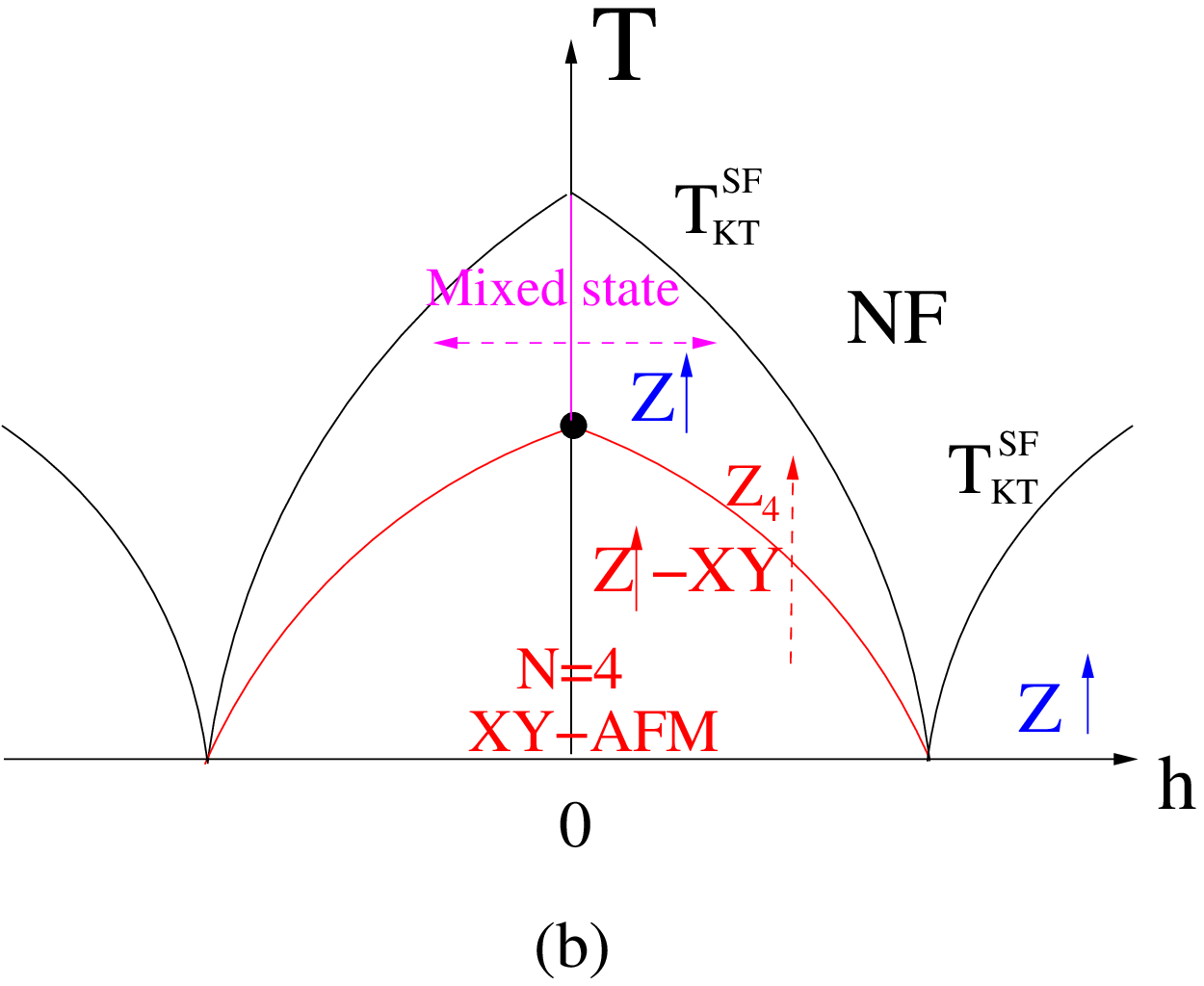}
    \caption{  The finite temperature phase transitions at a given $ t_s/t  $.
     The 1st order quantum phase transitions at $ T=0, h=0 $ are driven by roton dropping tuned by the Zeeman field.
     There is $ h \to -h $ symmetry of finite $ T_c $.
    (a) at a fixed $ 0 < t_s/t < \sqrt{2} $.  At $ h=0 $, there is a finite temperature  $ Z_4 $ clock melting transition
    with $ T_4 \sim \Delta_R $ from the $ N=2 $ XY-AFM ( a pure state )
    to a mixed state which consists of any mixture of spin up FM-SF at $ h > 0 $ and spin down FM-SF at $ h < 0 $.
    There is also a higher KT transition from the SF to a normal fluid (NF).
    (b) at a fixed $  t_s/t > \sqrt{2} $. There is  also a $ Z_4 $ clock transition above the
    $ N=4 $ XY-AFM to destroy the XY component of the spin with $ T_4 \sim \Delta_{3R} $. }
\label{rotondrive}
\end{figure}

 We also found that there are 3 gapped rotons near $(k_0,-k_0)$, $(-k_0,k_0)$, and $(-k_0,-k_0)$.
 In the  $ h \to 0$ limit, $ k_0 \to \pi/2 $,
 The two rotons $ \omega_{2}(\mathbf{q}) =\omega_{4}(\mathbf{q} ) $
 near $\mathbf{K}_2=(-\pi/2,\pi/2)$ and $ \mathbf{K}_4 =- \mathbf{K}_2 $
 start to touch down and also become two linear modes ( the upper dashed line in Fig.\ref{roton2gap}a ):
\begin{align}
    \omega_{2}(\mathbf{q}) = \sqrt{\frac{4n_0U}{4\sqrt{2}t_\text{s}+n_0U}
	    [t_\text{s}^2(q_x^2+q_y^2)-t^2(q_x-q_y)^2]}
\end{align}

 While the roton near $\mathbf{K}_3=(-\pi/2,-\pi/2)$ also starts to touch down, but become quadratic
 ( the lower dashed line in Fig.\ref{roton2gap}a ):
\begin{align}
    \omega_{3}( \mathbf{q} ) =
	\sqrt{a(q_x^4+q_y^4)+bq_xq_y(q_x^2+q_y^2)+cq_x^2q_y^2}
\end{align}
where
\begin{eqnarray}
	a & = & \frac{(t_\text{s}^2-t^2)[4(t_\text{s}^2-t^2)+\sqrt{2}n_0Ut_\text{s}]}
	       {2t_\text{s}(4t_\text{s}+\sqrt{2}n_0U)},~~ \nonumber  \\
	b & = & -\frac{t^2[8(t_\text{s}^2-t^2)+\sqrt{2}n_0Ut_\text{s}]}
	        {t_\text{s}(4t_\text{s}+\sqrt{2}n_0U)},~~   \nonumber  \\
	c & = & b+\frac{4(t^4+t_\text{s}^4)+\sqrt{2}n_0Ut_\text{s}^3}
	         {t_\text{s}(4t_\text{s}+\sqrt{2}n_0U)}
\end{eqnarray}

  It is easy to see the three linear modes $ \omega_{1} $ and $ \omega_{2} = \omega_{4}$
  are $ \sim \sqrt{ n_0 U t_s } $, but the quadratic mode $ \omega_{3} $  still exists even in the $ U=0 $ limit
  where it becomes identical to the free particle mode. This is clearly an arti-fact of the calculations
  at this order. In the following, we will show that the order from disorder analysis will open gaps
  to $ \omega_{2} = \omega_{4}$ and $ \omega_{3} $. So the transition is a first order one.


{\sl (c). The TT point approaching from the right. }

  If approaching the TT point from the right, taking
  $ t_s/t \to \sqrt{2} $ in Eq.\ref{gold1}, then $  \omega_{1} \sim \sqrt{ n_0 U t} |q_x-q_y| $ which becomes flat along the line $ q_x= q_y $.
  Of course, if we choose the PW state at the momentum $ \mathbf{K}_2$ in Eq.\ref{zxy},
  then $  \omega_{2} \sim  \sqrt{ n_0 U t} |q_x+q_y| $ which becomes flat along the line $ q_x= -q_y $.
  So the two flat directions from the left in Eq.\ref{cross} are just the combination of the one near $ (\pi/2, \pi/2 ) $ and
  another one near $ (\pi/2, -\pi/2 ) $.

{\bf 5. The Order from disorder phenomena and the $ N=4 $ XY-AFM at $ h=0, t_s/t > \sqrt{2} $. }


  We will study the nature of SF phase in the SOC dominated regime in Fig.\ref{phasedia}.
  In sharp contrast to the hopping dominated regime discussed above, the SOC term $ t_s $ shows its dominating
  effects even at the mean field level.
  But the effects of $ t $ drops off at the mean field level.
  There is a spurious $\tilde{\tilde{SU}}(2) $ symmetry at the 4 minima. The effects of $ t $ only show up in the quantum fluctuations.
  This is just opposite or dual to the hopping dominated regime in Sec.3 where the effects of $ t $ dominates, while $ t_s $ drops off
  at the mean field level as discussed above.

 At $ h=0 $, in the right of the T point $  t_s/t > \sqrt{2} $, $ k_0=\pi/2 $, the four minima are located at
 $\mathbf{K}_1=(\pi/2,\pi/2)$, $\mathbf{K}_2=(-\pi/2,\pi/2)$,
$\mathbf{K}_3=(-\pi/2,-\pi/2)$, and $\mathbf{K}_4=(\pi/2,-\pi/2)$. The corresponding four spinors in Eq.\ref{spinorn} become
\begin{align}
    \chi_1
	& =\frac{1}{\sqrt{2}}
	\begin{pmatrix}
	    -e^{-i\pi/4}\\
		1\\
	\end{pmatrix},\quad
    \chi_2
	=\frac{1}{\sqrt{2}}
	\begin{pmatrix}
	    e^{i\pi/4}\\
		1\\
	\end{pmatrix},\quad    \nonumber    \\
    \chi_3
	& =\frac{1}{\sqrt{2}}
	\begin{pmatrix}
	    e^{-i\pi/4}\\
		1\\
	\end{pmatrix},\quad
    \chi_4
	=\frac{1}{\sqrt{2}}
	\begin{pmatrix}
	    -e^{i\pi/4}\\
		1\\
	\end{pmatrix}\>.
\label{chi1234}
\end{align}

 The fact that the gapless mode at one momentum and
 the simultaneous  touching down of the three rotons at the other three momenta as $ h \rightarrow 0 $ indicates  all the four momenta
 $ (\pm \pi/2, \pm \pi/2 ) $ maybe involved in the ground at $ h=0 $. The most general mean field ground state can be written as:
\begin{eqnarray}
  \Phi_{h=0} = \sum_{i=1}^4  c_i e^{i \mathbf{K}_i \cdot \mathbf{r}_i} \chi_i
\label{fournodes}
\end{eqnarray}
 where $ | c_1 |^2 + |c_2|^2+ | c_3 |^2 + |c_4 |^2 =1 $.

  When evaluating its interaction energy,
  the most salient feature here is that  a loop term involving going around the 4 minima by one turn is a reciprocal lattice,
  so this Umklapp process is crucial even at the mean field level which is absent
  when $ h \neq 0 $ where $ k_0 \neq \pi/2 $ evaluated below Eq.\ref{fourPW}.
  Here the interaction energy takes a rather symmetric form:
\begin{align}
    E_{int}
	=\frac{N_sU}{2}\left( 1+\frac{1}{2} |Q|^2 \right)
\label{QQ}
\end{align}
  where the complex number $ Q=(\bar{c}_1+\bar{c}_3)(c_2+c_4)+i(c_1-c_3)(\bar{c}_2-\bar{c}_4) $.

Its minimization leads to  $ Q=0$.
By counting the number of free real parameters,
one can see the dimension of classically degenerate family ground state manifold is $8-1-2=5$.
In fact, as to be explained in the next subsection, in addition to the exact $ U(1)_c $ symmetry,
the Hamiltonian has a spurious $ SU(2)_s \times U(1) $ symmetry. So it counts as $ 5=1+3+1 $.
So the "quantum order from disorder" mechanism is needed to determine the quantum ground state
from the degenerate family Eq.\ref{fournodes}.
At the right Abelian point,  the spin $ SU(2)_s $ symmetry becomes exact, but the $ U(1) $ remains spurious.

Our previous work on the right Abelian point \cite{absf} showed that
the order from disorder mechanism on the spurious classically $ U(1) $ degenerate family  will pick up
a plane-wave state ( namely,  one of the $ \chi_i, i=1,2,3,4 $ in Eq.\ref{chi1234} )
plus its spin $ SU(2)_s $ related family in $t/t_s\to0$ limit.
For a small deviation from the Abelian limit, i.e. $0<t/t_s\ll1$, the exact spin $ SU(2)_s $ symmetry
becomes spurious at mean field level. So we expect the order from disorder mechanism will pick up
a true quantum ground state from this  spin $ SU(2)_s $ related family.
We did the order from disorder analysis in the three channels generated by the three $ SU(2) $
generators $  R_x^{soc}( \phi )=e^{i \phi \sum_i (-1)^{i_y}\sigma_x},
R_y^{soc}( \phi )=e^{i \phi \sum_i (-1)^{i_x}\sigma_y}, R_z^{soc}( \phi )=e^{i \phi \sum_i (-1)^{i_x+i_y}\sigma_z} $
which generate the orbital orders $ (\pi,0), ( 0, \pi) $ and $ ( \pi, \pi) $ respectively.
Due to the $ [ C_4 \times C_4]_D $ symmetry $ (\pi,0) $ and $ ( 0, \pi) $ channels are identical, so
one only need to focus on$ (\pi,0) $. The results are shown in Fig.\ref{k1k3fig}.
The details are given in the Method section where we find the quantum ground state happens in the
  $ \mathbf{K}_1-\mathbf{K}_3=( \pi, \pi) $ channel:
\begin{align}
    \Psi_{0,R1}
	=\frac{1}{2}\left[
	    \begin{pmatrix}
		-e^{-i\pi/4}\\
		    1\\
	    \end{pmatrix}
	    e^{i\mathbf{K}_1\cdot\mathbf{r}_i}
	    \pm i
	    \begin{pmatrix}
		e^{-i\pi/4}\\
		    1\\
	    \end{pmatrix}
	    e^{i\mathbf{K}_3\cdot\mathbf{r}_i}
	\right]
\label{k1k30}
\end{align}
which corresponds to $ (c_1,c_2,c_3,c_4)=1/\sqrt{2} (1,0, \pm i,0) $ in Eq.\ref{fournodes}.
Its corresponding spin-orbit structure is:
\begin{align}
    \mathbf{S}_i=\langle \frac{\hbar}{2}\vec{\sigma}\rangle
    =\pm \frac{\hbar}{2}\big( (-1)^{x+y}/\sqrt{2}, (-1)^{x+y+1}/\sqrt{2},0 \big)
\label{AFMXY2}
\end{align}
  which is only the $ (+,-) $ or $ (-,+) $ combinations in Eq.\ref{AFMXY1},
  so it only has the $ d_{R1}=2 $ degeneracy \cite{rotation11}.

  Equivalently,  it can happen in the $ \mathbf{K}_2-\mathbf{K}_4=( \pi, \pi) $ channel:
\begin{align}
    \Psi_{0,R2}
	=\frac{1}{2}\left[
	    \begin{pmatrix}
		e^{i\pi/4}\\
		    1\\
	    \end{pmatrix}
	    e^{i\mathbf{K}_2\cdot\mathbf{r}_i}
	    \pm i
	    \begin{pmatrix}
		-e^{i\pi/4}\\
		    1\\
	    \end{pmatrix}
	    e^{i\mathbf{K}_4\cdot\mathbf{r}_i}
	\right]
\label{k2k40}
\end{align}
which corresponds to $ (c_1,c_2,c_3,c_4)=1/\sqrt{2} (0,1,0, \pm i ) $ in Eq.\ref{fournodes} with $ (+,+) $ or $ (-,-) $
combinations in Eq.\ref{AFMXY1}. It only has the $ d_{R2}=2 $ degeneracy.

When combining Eq.\ref{k1k30} with Eq.\ref{k2k40}, it is a 4-fold degenerate state related by the $ [C_4 \times C_4]_D $  symmetry.
Surprisingly, the combined spin-orbital structure of the two sets of Bose condensations is the same as that on the right Eq.\ref{AFMXY1}
with the same degeneracy $ d=d_{R1} + d_{R2} =4 $.
Even so, the boson wavefunctions Eq.\ref{k1k30} and Eq.\ref{k2k40} on the right with $ N=4 $ condensation momenta
differ from Eq.\ref{deg} with $ N=2 $ condensation momenta on the left \cite{rotation11},
so this must be a TPT  which resembles the TPT in a fermionic system \cite{topo1,topo2}.
This could be the first bosonic analog of the fermionic topological Lifshitz  transition
of free fermions with Fermi surfaces, Dirac points or Weyl points \cite{topo1,topo2}.


We also computed the
4 roton gaps $ \Delta_{2R}= \Delta_{4R} \gg \Delta_{3R}  $ generated from the  " quantum order from disorder "
shown in Fig.\ref{roton2gap}.

 {\bf 6. The physical picture near the left Abelian point $ (h=0, t_s=0 ) $ and near the right Abelian point $ (h=0, t=0 ) $ }

 At the left Abelian point $ (h=0, t_s=0 ) $, after the transformation $ a_{i\downarrow}\to(-1)^ia_{i\downarrow}$,
 the spin $ SU(2) $ symmetry becomes evident in the rotated basis $\tilde{\mathbf{S}}_i=((-1)^{i_x+i_y}S_i^x,(-1)^{i_x+i_y}S_i^y,S_i^z)$.
 This left Abelian point is controlled by the spin-conserved hopping term $ t $.
 There is no coupling between the off-diagonal long range order
 in the SF sector and the magnetic order in the spin order. The breaking of $ U(1)_c $ and that of spin $ SU(2) \rightarrow U(1) $
 leads to the linear superfluid Goldstone mode and  the quadratic FM spin wave mode $ \omega \sim k^2 $ respectively and separately.

 When slightly away from the left Abelian point along the horizontal axis, a small SOC  $ t_s $ term will break the $ SU(2) $ symmetry. However, at the minimum $ (0,0) $ and $ (\pi,\pi) $, the $ t_s $ term still drops out, so the SOC still plays no roles at the mean field level, however, it still plays important roles in the quantum fluctuations. So the exact $ SU(2) $ symmetry becomes a spurious one, so an order from disorder
 phenomena is needed to find the quantum ground state from the classically degenerate family of states Eq.\ref{deg} due to this spurious $ SU(2) $ symmetry. Indeed, as shown in the order from disorder analysis in the Sec.3, it is quantum fluctuations induced by the SOC term $ t_s $ which picks up the $ N=2 $ XY AFM state as the quantum ground state, also open the gap  $ \Delta_R $ to the roton mode shown in Fig.\ref{roton1gap}.
 At $ t_s =0 $, the $ N=2 $ XY-AFM becomes a FM in the XY plane in the $\tilde{\mathbf{S}} $ basis,
 so the roton mode is nothing but the FM spin wave mode.
 The $ t_s $ term opens  a gap to the FM spin wave mode and transfer it to a pseudo-Goldstone mode.
 As shown in the method section, in the  $\tilde{\mathbf{S}} $ basis,  the second order perturbation in $ t_s $ picks up the XY plane,
 the fourth order perturbation selects $ \phi=\pi/4 $ and the gap $ \Delta_R \sim t^{3}_s $.
 Transforming back to the original basis, it is the $ N=2 $ XY-AFM state.
 So the result is consistent with the order from disorder analysis done in the original basis in Sec.3.

 When slightly away from the left Abelian point along the vertical axis, the state is just the Z FM state, the Zeeman field simply opens a gap to
 the FM spin wave mode and  transfer it to a pseudo-Goldstone mode.

 In the right Abelian point $ (h=0, t=0 ) $, the spin $ SU(2) $ symmetry becomes evident
 in the rotated basis  $\tilde{\tilde{\mathbf{S}}}_i=((-1)^{i_y}S_i^x,(-1)^{i_x}S_i^y,(-1)^{i_x+i_y}S_i^z)$
 which was used in \cite{rh,rhht,rhh},
 This right Abelian point is controlled by the SOC term $ t_s $.
 As shown in \cite{absf}, there is a strong quantum fluctuations induced by the coupling between the off-diagonal long range order
 in the SF sector and the magnetic order in the spin order ( This is in sharp contrast to
 the physics at the left Abelian point discussed above where there is no such coupling.)
 The complete symmetry breaking $ U(1)_c \times SU(2) \rightarrow 1 $ leads to
 4 linear gapless modes. When transforming the Abelian gauge to the present non-abelian gauge with $ ( \alpha=\pi/2, \beta=\pi/2 ) $,
 the $ 90^{\circ} $ co-planar state becomes the $ N=4 $ XY-AFM state,
 the 4 linear modes are shifted to the 4 minima $ (\pm \pi/2, \pm \pi/2) $.

 When slightly away from the right Abelian point along the horizontal axis, a small $ t $ term will break the $ SU(2) $ symmetry.
 However, the $ t $ term still drops out at the 4 minima $ ( \pm \pi/2, \pm \pi/2 ) $, it
 still plays no roles at the mean field level.
 So the exact spin $ SU(2) $ symmetry becomes a spurious one, so an order from disorder
 phenomena is needed to find the quantum ground state from the classically degenerate family of states Eq.\ref{fournodes} due to this spurious $ SU(2) $ symmetry. So we expect one gapless mode remains, the other three become gapped pseudo-Goldstone modes.
 Indeed, as summarized   in the order from disorder analysis in the Sec.5 and detailed in the method section, the small $ t $ term
 picks up the $ N=4 $ XY AFM state Eqs.\ref{k1k30},\ref{k2k40}
 as the quantum ground state, also open the gap  $ \Delta_{2R}= \Delta_{4R} $ to the two linear modes
 at $ \mathbf{K}_2=(-\pi/2, \pi/2), \mathbf{K}_4= - \mathbf{K}_2 $ and  $ \Delta_{3R} \ll \Delta_{2R} $ to the quadratic mode
 at $ \mathbf{K}_3=(-\pi/2, -\pi/2) $. Of course, the linear gapless SF Goldstone
 mode at $  \mathbf{K}_1=(\pi/2, \pi/2) $ remains due to the $ U(1)_c $ symmetry breaking.

 When slightly away from the right Abelian point along the vertical axis, there should be two gapless Goldstone modes
 one at $ (0,0) $ due to  breaking the $ U(1)_c $, another at $ (\pi,\pi) $ due to  breaking the $ U(1)_s $ which is a subgroup
 of $ SU(2)_s $. Indeed, taking the results on Z-XY state in Sec.4, as $ \beta \rightarrow \pi/2^{-} $ at $ h \neq 0 $,
 namely, approaching the right axis at $ h \neq 0 $, we find $ \Delta_2=\Delta_4 $ and $ \Delta_3 $ all approach zero.
 However, the order from disorder analysis at the right axis \cite{junsen} opens the gap to the $ ( \pi,0 $ and $ (0,\pi) $ channel.
 While the gapless mode at $ ( \pi,\pi) $ is the just the Goldstone mode due to the $ U(1)_s $ breaking.

 So all the quantum phase transitions in Fig.1 can be considered as the one from the phases
 controlled by the conventional hopping $ t_s/t=0 $ ( Abelian regime )  to the phases
 controlled by the SOC term $ t/t_s=0 $ ( Non-Abelian regime )
  ( Note that $ t/t_s=0 $ is the $ \pi $ flux  Abelian point.)

{\bf 7. Finite Temperature Phase transitions. }

\begin{figure}[!htb]
\centering
\includegraphics[width=0.22\textwidth]{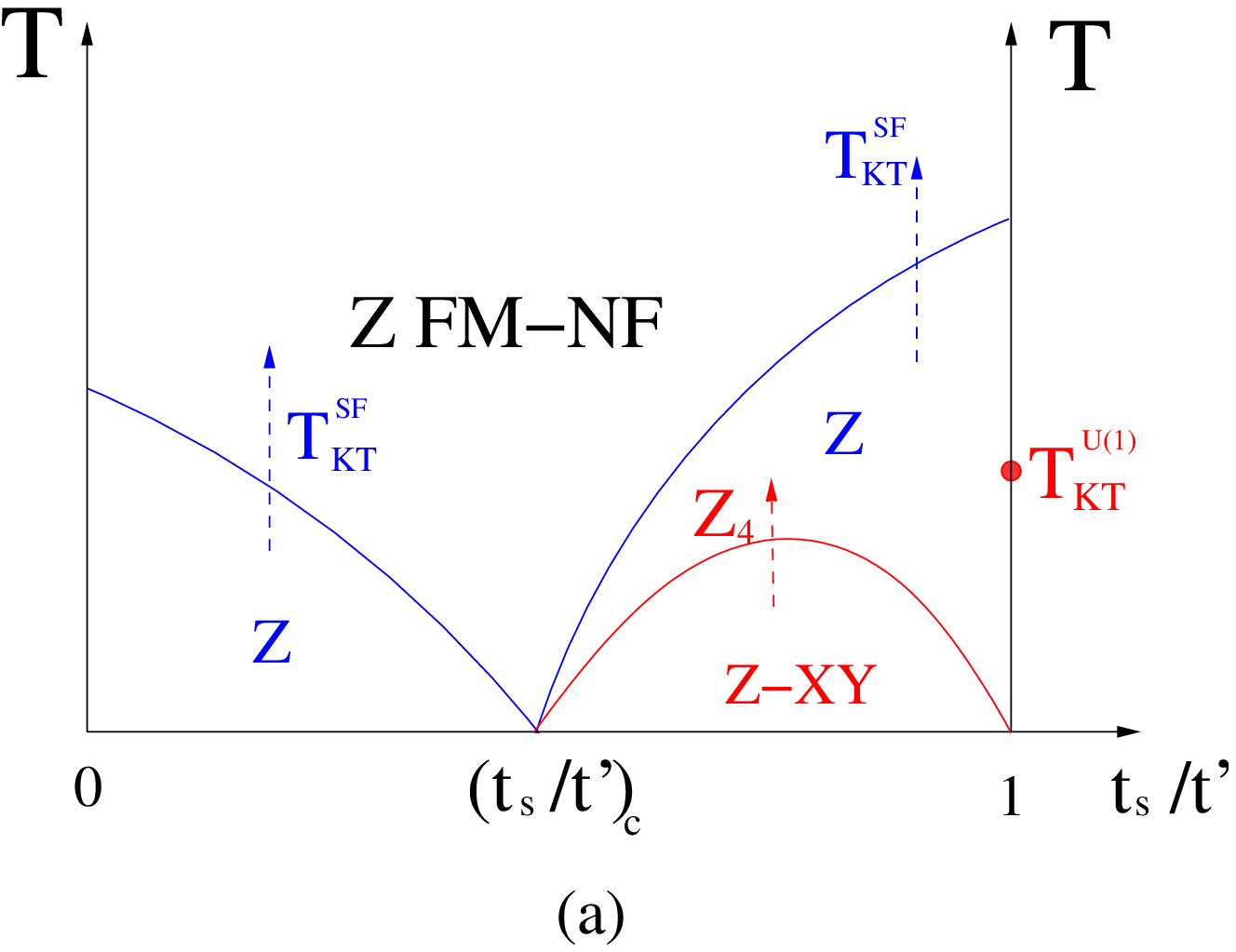}
\includegraphics[width=0.22\textwidth]{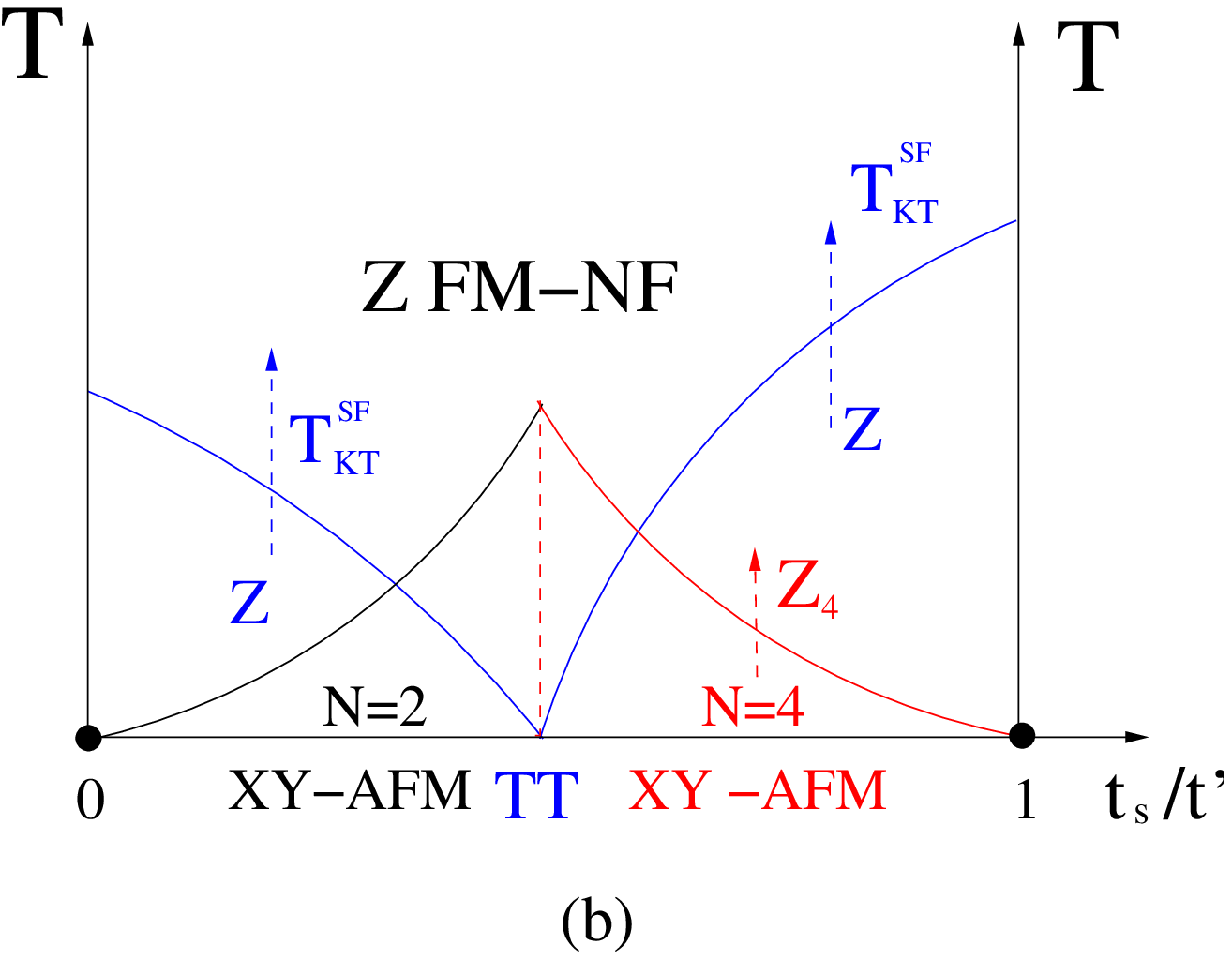}
    \caption{ The finite temperature phase transitions at a given $ h $ with $ t^{\prime}= \sqrt{t^2 + t^2_s} $.
     (a) at a fixed $ h \neq 0 $.
      The quantum phase transitions at $ h>0, T=0 $ is a Bosonic Lifshitz transition with the dynamic exponent $ z_x=z_y=2 $
     driven by the softening of superfluid Goldstone mode.
     There are KT transitions at $ T^{SF}_{KT}  $ from the SF to a normal fluid (NF) on both sides.
     There is lower $ Z_4 $ clock melting transition to destroy the  XY component of the spin on the right side $ t_s/t^{\prime} > ( t_s/t^{\prime} )_c $.
     On the right axis, there is also a KT transition due to the $ U(1) $ symmetry breaking in the $\tilde{\tilde{SU}}(2) $ basis.
     (b)  at a fixed $ h = 0 $.
     The Topological Tri-critical (TT) point at $ h=0, t_s/t^{\prime}=\sqrt{2/3}, T=0 $.
     There are $ Z_4 $ clock transition with $ T_4 \sim \Delta_R $ above the $ N=2 $ XY-AFM in Fig.6b and
     $ T_4 \sim \Delta_{3R} $ above the $ N=4 $ XY-AFM in Fig.7b respectively.
     The left and right dot stands for the  the $\tilde{SU}(2) $ and  $\tilde{\tilde{SU}}(2) $  symmetry respectively.  }
\label{sfdrive}
\end{figure}

  Now we briefly discuss the finite temperature phases and phase transitions above all the SFs in Fig.1.
  Of course, due to the $ U(1)_c $ symmetry breaking in the SF phase, there is always a KT transition $ T^{SF}_{KT} $ above all the SFs.
  Due to the correlated spin-bond orders of the SFs, there are also other phase transitions associated  with the restorations of these
  orders at  finite temperatures shown in Fig.\ref{rotondrive} and Fig.\ref{sfdrive}. The nature of these finite temperature transitions can be qualitatively determined by the degeneracy of the ground states:
  Z FM-SF (spin up or down ) ( $ d=1 $ ), Z-XY FM-SF (spin up or down ) ( $ d=4 $ ),  $ N=2 $ and $ N=4 $  XY-AFM ( $ d=4 $ )
  which breaks not only the global $ U(1)_c $ symmetry, but also the spin-orbital $ [C_4 \times C_4]_D $ symmetry.
  Especially, at $ h=0 $, there is a finite temperature $ Z_4 $ clock transition from  both the $ N=2 $ and $ N=4 $
  XY-AFM to a mixed state which is
  a mixture of the phases at the two sides $ h >0 $ and $ h < 0 $ at any ratios in Fig.\ref{rotondrive}.
  If the $ T=0 $ transition is the first order one  driven by roton dropping tuned by the Zeeman field
  as in Fig.\ref{rotondrive}, the SF KT transition is higher near the transition.
  If the $ T=0 $ transition is the second order bosonic Lifshitz one driven by the softening
  of the SF Goldstone mode tuned by $ t_s/t $ as in Fig.\ref{sfdrive}, the SF KT transition is lower  near the transition.
  But the two finite phase transition lines could cross when away from the $ T=0 $ transition.
  The gap $ \Delta_R $ above the $ N=2 $ XY-AFM in Fig.6b and $\Delta_{3R} $ above the $ N=4 $ XY-AFM in Fig.7b
  directly determine the finite $ T_c $ of the two $ Z_4 $ clock transitions above the two XY-AFM states respectively in Fig.9b.


{\bf 8.  Implications on ongoing experimental efforts on QAH $^{87}$Rb  atoms to detect the many body phenomena  }

  In the ongoing experiment at USTC \cite{2dsocbec}, the BEC  has $ N \sim 3 \times 10^5 $ atoms and trapped within the diameter $ d= 80 \mu m $,
  so one lattice site has about $ n= 10 $ atoms.
The short-range Hubbard like interaction $ U= \frac{ 4 \pi \hbar^2 a_s}{m} $  in Eq.\ref{qah}.
Plugging in the S-wave scattering length of the $ ^{87}$Rb  atoms $ a_s=103 a_0 $ where $ a_0 $ is the Bohr radius
and the mass of the Spinor bosons $ ^{87}$Rb  atoms lead to $ U \sim 50 nK $.
The interaction also has a negligible spin anisotropy with $ \lambda \sim 1 $.

   It is easy to evaluate the density-spin correlation functions whose poles should be determined by
   the Bogliubov spectrum in Fig.\ref{roton1},\ref{GR},\ref{roton1gap} with the corresponding spectral weight. So all interesting behaviors of the Goldstone mode near $ (0,0) $ and the roton mode near $ (\pi,\pi) $ in these figures can be precisely monitored and  mapped out by the Bragg spectroscopies \cite{becbragg} used in the same lab before. For example,
  taking a point such as $ (t_s, h)=(1,1) $ landing in the $ (0,0) $ spin up FM SF in Fig.1,
  Tuning $ h \rightarrow 0 $ to observe the roton softening by the Bragg spectroscopies \cite{becbragg} is  a easy task.
  However approaching the bosonic Lifshitz transition boundary in Fig.1 is more challenging.
  The hopping $ t $ is determined by the depth of the optical lattice generated by a  standing  wave laser,
  while the SOC $ t_s $ is determined by the strength of the traveling Raman laser.
  But the numerical value of the ratio $ t_s/t $ was not estimated in the experiment\cite{2dsocbec}, because it is not
  important to the single particle band structure anyway which shows QAH as long as the SOC strength $ t_s \neq 0 $.
  However, as shown in Fig.1, it is important in
  the many body phenomena explored here. So increasing the depth of the optical lattice and also the strength of the Raman laser
  should be able to increase the ratio $ t_s/t > \sqrt{2} $ to reach SOC dominated non-Abelian regime in Fig.1.
  Then the novel many body phenomena of this non-Abelian regime can be experimentally detected.

   The Time of Flight (TOF) image after a time $ t $ is given by \cite{blochrmp}:
 \begin{equation}
     n( \vec{x} )= ( M/\hbar t )^3 f(\vec{k}) G(\vec{k})
 \end{equation}
   where $ \vec{k}= M \vec{x}/\hbar t $,  $ f(\vec{k})= | w(\vec{k}) |^2 $ is the form factor due to the
   Wannier state of the lowest Bloch band of the optical lattice and
   $ G(\vec{k}) = \frac{1}{N_s} \sum_{i,j} e^{- \vec{k} \cdot ( \vec{r}_i- \vec{r}_j ) }
    \langle \Psi^{\dagger}_i \Psi_j \rangle $ is the equal time boson structure factor.
   For small condensate depletion in the weak coupling limit $ U/t \ll 1 $,
   $ \langle \Psi^{\dagger}_i \Psi_j \rangle \sim \langle \Psi^{\dagger}_{0i} \Psi_{0j}  \rangle$
   where $ \Psi^{\dagger}_{0i} $ is the condensate wavefunction.
   Obviously, the wavefunctions of the PW state leading to
   Z FM-SF (spin up or down ) ( $ d=1 $ ), the PW state Eq.\ref{zxy} leading to the  Z-XY FM-SF (spin up or down ) ( $ d=4 $ )
   can be easily detected. At $ h=0 $, due to the order from disorder mechanism,
   both the  $ N=2 $ XY-AFM boson wavefunction Eq.\ref{deg} in the hopping dominated Abelian regime and
   the$ N=4 $ XY-AFM boson wavefunctions  Eq.\ref{k1k30}, Eq.\ref{k2k40} in the SOC dominated non-Abelian regime lead to
   the spin-orbital XY-AFM ( with the degeneracy $ d=4 $ ) in Eq.\ref{AFMXY1}.
   The TOF can detect the differences in the BEC condensation topology
   between the two bosonic wavefunctions which lead to the same set of spin-orbital structures.
   So the TOF can detect all the quantum and topological phenomena in Fig.1.

 The spin-orbital structures, the excitations above the SF phases across the whole BZ,
 the transitions driven by the roton dropping or the softening of the SF Goldstone bosons and
 the location of the TT point can be precisely determined by various experimental techniques
 such as dynamic or elastic, energy or momentum resolved, longitudinal or
 transverse Bragg spectroscopies \cite{becbragg,bragg1,bragg2},
 specific heat measurements \cite{heat1,heat2}  and {\sl In-Situ} measurements \cite{dosexp}.

{\bf  DISCUSSIONS }

    The TPT in non-interacting fermions can be characterized by the change in the topology of Fermi Surface (FS),
    the Dirac point  or Wely point \cite{topo1,topo2} where the low energy fermionic excitation stay.
    Here, for bosonic SF system which is necessarily an interacting system, we propose to characterize the
    TPT by the change in the topology of BEC condensation momenta where the low energy bosonic excitation stay.
    The TPT is protected by a anti-unitary discrete symmetry which is the $ Z_2 $ reflection symmetry in the present context.
    Under the $ Z_2 $ transformation, the momentum changes as
    $ \vec{k} \rightarrow -\vec{k}+ ( \pi,\pi) $.
    So the  $ N=2 $ BEC condensation momenta in the left of Fig.\ref{tpt} transform to each other,
    while the  $ N=4 $ BEC condensation momenta in the right of Fig.\ref{tpt} are invariant under the $ Z_2 $ transformation.
    This is the only two possibilities which lead to the ground states with
    the XY-AFM spin-orbital structure  respecting the $ Z_2 $ symmetry.

    The "order from quantum disorder" phenomena and the associated mass generations
    have been studied in the geometrically frustrated magnets \cite{ringrev,sachdev}.
    In this work, for the first time, we studied
    the novel frustrated phenomena due to the SOC leading to QAH at zero Zeeman field in Fig.1.
    Most interestingly, the quantum ground $ N=2 $ and $ N=4 $ XY-AFM state  selected by the "order from quantum disorder" phenomena
    is a pure state instead of a mixed state of the two states at $ h=0^{+} $ and $ h=0^{-} $.
    This is in sharp contrast to the case in the Rashba SOC induced frustration in quantum spin systems \cite{rhrashba}.
    These studies  in the SOC frustrated superfluids
    should open new horizons to its original discovery in the context of frustrated magnetisms.

     It was known that Haldane model \cite{haldane0} in a honeycomb lattice is the first QAH model.
     Due to the two sublattice structure of the honeycomb lattice, it only needs Abelian gauge field.
     The effects of interaction on the Haldane model in its spin-less or spinful version has been
     investigated by various groups \cite{canada}.  However, to realize QAH in a square lattice, a non-abelian gauge field
     ( or SOC  ) $ t_s $ term in Eq.\ref{qah} is needed.  Surprisingly, the interaction effects of QAH model in a square lattice
     has been completely overlooked so far. Obviously, due to its Non-Abelian nature,
     the many body phenomena are quite different than those in the Haldane-Hubbard model.



    If reversing the sign of the down hopping term in Eq.\ref{qah},  then it maps to the
    isotropic Rashba model $ \alpha=\beta $  studied in Ref.\cite{rh,rhrashba}
    with $ \cos \alpha= t/\sqrt{t^2+t^2_s} $ in a Zeeman field $ H $.
    Of course, it is this sign difference which introduces the third Pauli spin $ \sigma_z $ even
    at the zero Zeeman field and introduces topological band structure at a finite Zeeman field
    in the QAH model. Therefore, the QAH model displays dramatically different many body phenomena
    than the interacting Rashba or Weyl SOC systems \cite{rh,rhht,rhh,rhrashba,rasf}.
    So the two systems are  complementary and shed lights to each other.
    It was known that the  Rashba SOC band structure does not have topological properties.
    It remains interesting to explore how this (non)-topological  band structure affects
    the many body phenomena for spinor bosons at both  weak and strong interaction.

{\bf  Methods }

{\bf 1. Perturbation theory at a small $ t_s/t $ to calculate the roton gap in the
 $ N=2 $ XY-AFM phase in the hopping dominated Abelian regime. }

 In the left side of Fig.1, at small $ t_s/t $ and $ h=0 $, we perform the perturbation calculation
 in $ t_s $ which is independent, but complementary to the order from disorder analysis done in Sec.3.
 We work in $\widetilde{SU}(2)$ basis in Fig.1.
 When evaluating the ground-state energy upto fourth order in $ t_s $, we determine the state
 with $\theta=\pi/2$ and $\phi=\pi/4$ to be the round state, which, after transforming back
 the original basis, is the $ N=2 $ XY-AFM state on the left in Fig.1.

 We begin with an unperturbed Hamiltonian in the $\widetilde{SU}(2)$ basis with no SOC  $ t_s=0 $,
\begin{align}
	H_0=-t\sum_{\langle ij\rangle\sigma} a_{i\sigma}^\dagger a_{j\sigma}
	+\frac{U}{2}\sum_i n_{i}(n_{i}-1)-\mu\sum_i n_{i}
\end{align}
where $n_{i\sigma}=a_{i\sigma}^\dagger a_{i\sigma}$
and $n_i=n_{i\uparrow}+n_{i\downarrow}$. It is essentially a spinor boson Hubbard model.
Adding $ \pi $ flux to this model leads to the Abelian point on the right in the $ \widetilde{\widetilde{SU}}(2)$ basis \cite{absf}.

 Because it has the $ \widetilde{SU}(2)_s $ symmetry,
 one can condense the spinor bosons at $ \bf{k}=0 $ with the spin symmetry breaking direction
 along $ (\theta, \phi) $. The $ H_0 $  can be diagonalized by a $ 4 \times 4 $ Bogoliubov transformation:
\begin{align}
    H_0=E_0+
	\sum_\mathbf{k}(\omega_{1,\mathbf{k}}\alpha_{\mathbf{k}}^\dagger \alpha_{\mathbf{k}}
	    +\omega_{2,\mathbf{k}}\beta_{\mathbf{k}}^\dagger \beta_{\mathbf{k}})
\label{h0left}
\end{align}
where $\omega_{1,\mathbf{k}}=4t-2t(\cos k_x+\cos k_y)\sim t(k_x^2+k_y^2)$ is the Ferromagnetic (FM) spin mode
and $\omega_{2,\mathbf{k}}=\sqrt{\omega_{1,\mathbf{k}}(\omega_{1,\mathbf{k}}+2n_0U)}\sim\sqrt{2n_0Ut(k_x^2+k_y^2)}$
is the superfluid (SF) Goldstone mode which, in fact, is the same as that in a single component case.
In sharp contrast to the $ \pi $ flux Abelian point in the right of Fig.1 where there is a coupling between the
density mode and spin mode \cite{absf}, here the FM spin mode is decoupled from the SF density mode.

  In the $\widetilde{SU}(2) $ basis, one can write the SOC term as:
\begin{align}
    H_\text{s}
	=2t_\text{s}\sum_k[\gamma_\mathbf{k} a_{\mathbf{k}\uparrow}^\dagger a_{\mathbf{k+Q}\downarrow}+h.c.]
\end{align}
  where  $ \gamma_\mathbf{k}=\sin k_x-i\sin k_y $.

  Using the fact that the unperturbed ground-state $|0\rangle$ is the vacuum of
  $\alpha_\mathbf{k}$ and $\beta_\mathbf{k}$ in Eq.\ref{h0left}:
  $\alpha_\mathbf{k}|0\rangle=\beta_\mathbf{k}|0\rangle=0$, One can just apply non-degenerate perturbation
  theory in $ H_\text{s} $ to calculate the corrections to the ground state energy.

  Because $\langle 0|H_\text{s}|0\rangle=0$, one can see that odd order corrections vanish, so
\begin{align}
    \delta E=\delta E^{(2)}+\delta E^{(4)}+\cdots
\end{align}
where
\begin{align}
    \delta E^{(2)}&=\langle 0|H_\text{so} g H_\text{so}|0\rangle,~~~g=\sum_{n\neq0}\frac{|n\rangle\langle n|}{-E_n}  \nonumber  \\
    \delta E^{(4)}&=
	\langle 0|H_\text{s}gH_\text{s}gH_\text{s}gH_\text{s}|0\rangle
	-\langle0|H_\text{s}gH_\text{s}|0\rangle
	 \langle0|H_\text{s}g^2H_\text{s}|0\rangle
\end{align}

  After some length, but straightforward manipulations, we find:
\begin{align}
	\delta E^{(2)}=-t_\text{s}^2[a-b\cos2\theta],
\label{e2}
\end{align}
where $ a, b >0 $. so $\delta E^{(2)}$  reaches its minimum at $\cos2\theta=-1$ leading to $\theta=\pi/2$.

Unfortunately, $ \delta E^{(2)} $ is independent of $\phi$, so one must work out upto the fourth order corrections.
After some length, but straightforward manipulations, we find that at fixed $ \theta= \pi/2 $:
\begin{align}
	\delta E^{(4)}=-t_\text{s}^4[c-d \cos 4\phi],
\label{e4}
\end{align}
where $ c, d >0 $. So $\delta E^{(4)}$  reaches its minimum at $\cos4\phi=-1$ leading to  $\phi=\pi/4$.

\begin{figure}[!htb]
\centering
    \includegraphics[width=0.90\linewidth]{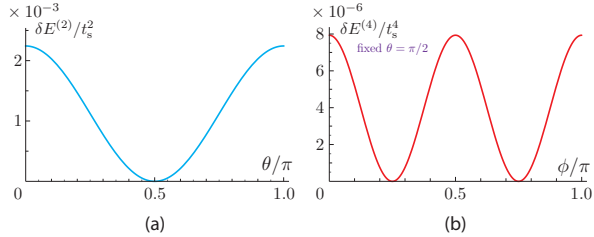}
    \caption{(a) The 2nd order correction to the ground-state energy
	$\delta E^{(2)}$ as a function of $\theta$ with fixed $ \phi=\pi/4 $ and $n_0U=t=1$.
	(b) The 4th order correction to the ground-state energy
	$\delta E^{(4)}$ as a function of $\phi$ with fixed $\theta=\pi/2$ and $n_0U=t=1$.
	In order to see the energy difference clear, we shifted all the energy by its minimum.}
\label{pert24}
\end{figure}

\begin{figure}[!htb]
\centering
\includegraphics[width=\linewidth]{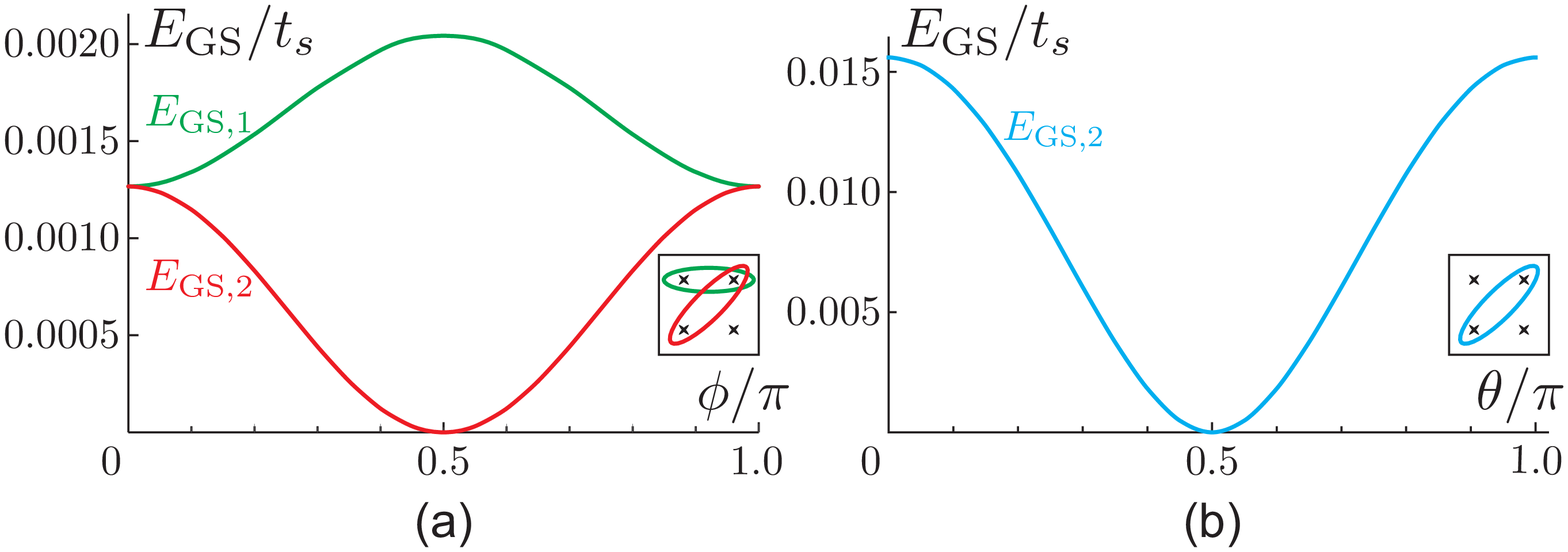}
\caption{
    (a) The quantum fluctuations corrected ground state energy $E_\text{GS}$ as a function of $ \phi $ at
    fixed $t/t_s=1/3$ and $ \theta=\pi/2 $, and $n_0U/t_s=1$.
	$E_\text{GS,1}$ is the ground-state energy in the $ (\pi,0) $ channel consisting of the condensations
    at $\mathbf{K}_1$ and $\mathbf{K}_2$ in Eq.\ref{k1k2}.
	$E_\text{GS,2}$ is the ground-state energy in the $ (\pi,\pi) $ channel consisting of the condensations
    at $\mathbf{K}_1$ and $\mathbf{K}_3$ in Eq.\ref{k1k3}.
	(b)
	$E_\text{GS,2}$ as a function of $ \theta $ at a fixed $ \phi=\pm \pi/2 $.
	In order to see the energy difference clear, we shifted all the energy by its minimum at $E_\text{GS,2}(\pi/4)$.  }
\label{k1k3fig}
\end{figure}

In Fig.\ref{pert24}, we plot the numerical values of $\delta E^{(2)}$ and $\delta E^{(4)}$
which match the numerical data from the direct calculations. Fig.\ref{pert24} is consistent with
Fig.\ref{ABfigure}, so we reach the same ground state $ N=2 $ XY-AFM on the left side of Fig.1 from
two independent methods. From Eq.\ref{e2} and \ref{e4}, one can extract $ A \sim t^2_s, B \sim t^4_s $, so
Eq.\ref{deltar} leads to $ \Delta_R \sim t^{3}_s $ shown in Fig.\ref{roton1gap}.



 In principle, one can apply a similar perturbation theory near the right $ \pi $ flux Abelian point in Fig.1 where $ t_s \neq 0, t=0 $.
 In the $\tilde{\tilde{\mathbf{S}}} $ basis,
 one may treat $ t $ as a small perturbation when slightly away from the   $ \pi $ flux Abelian point.
 Unfortunately, in practice, due to the coupling between the SF mode and the spin mode already at the   $ \pi $ flux Abelian point,
 the perturbation calculations are too complicated to extract useful information.
 It seems the " order from quantum disorder" analysis is the only
 practical way to determine the true ground state and also calculate the spectrum which we do in the following.


{\bf 2. Calculating the roton gaps in the $ N=4 $ XY-AFM phase in the SOC dominated Non-Abelian regime by order from disorder analysis }

 As shown in the section 6, the exact symmetry at the right Abelian point in the rotated basis $\tilde{\tilde{\mathbf{S}}}_i=((-1)^{i_y}S_i^x,(-1)^{i_x}S_i^y,(-1)^{i_x+i_y}S_i^z)$  becomes a
 spurious one when away from it by $ t > 0 $. So  order from disorder analysis is needed to calculate the roton gaps
 $\Delta_{2R}= \Delta_{4R} $ at $ (\pi,0) $ or $ ( 0, \pi) $ and $ \Delta_{3R} $ at $ ( \pi,\pi) $ in the $ N=4 $ XY-AFM phase.

{\sl (a). Estimations of $\Delta_{2R}= \Delta_{4R}$. }

The Spin-orbital rotation $R_y^{soc}( \phi )=e^{i \phi \sum_i (-1)^{i_x}\sigma_y}$ acting on the $ \chi_1 $ state
generates a superposition of condensations at $\mathbf{K}_1$ and $\mathbf{K}_2$.
The most general state in such a superposition is:
\begin{align}
    \Psi^{0}_{12}=c_1\chi_1e^{-i\mathbf{K}_1\cdot\mathbf{r}_i}+c_2\chi_2e^{-i\mathbf{K}_2\cdot\mathbf{r}_i}
\label{k1k2}
\end{align}
where one may note that $ \chi_1, \chi_2 $ are not orthogonal as listed in Eq.\ref{chi1234}.
Of course, due to the $ [C_4 \times C_4 ]_D $ symmetry, $\mathbf{K}_1$ and $\mathbf{K}_4$ work equally well
with $ \Psi^{0}_{14} $ in Eq.\ref{k1k2}. In the following, we will work in this sub-manifold to estimate the gap $\Delta_{2R}= \Delta_{4R} $.

   Using $ \chi_1 $ and $ \chi_2 $, we can construct the two eigenstates of $ Q_y= \sum_i (-1)^{i_x}\sigma_y $ as
\begin{align}
    \Psi_{\text{Y-x},\pm}
	=\frac{1}{\sqrt{2}}
	\left[\chi_1e^{-i\mathbf{K}_1\cdot\mathbf{r}_i}
	\pm  e^{i\pi/4}\chi_2e^{-i\mathbf{K}_2\cdot\mathbf{r}_i}\right]
\end{align}
 which can be used to re-parameterize Eq.\ref{k1k2} as:
\begin{align}
    \Psi^{0}_{12} =e^{i\phi/2}\cos(\theta/2)\Psi_{\text{Y-x},+}+e^{-i\phi/2}\sin(\theta/2)\Psi_{\text{Y-x},-}
\end{align}
where
\begin{align}
    c_1 & =[e^{i\phi/2}\cos(\theta/2)+e^{-i\phi/2}\sin(\theta/2)]/\sqrt{2},\quad     \nonumber  \\
    c_2 & =e^{i\pi/4}[e^{i\phi/2}\cos(\theta/2)-e^{-i\phi/2}\sin(\theta/2)]/\sqrt{2}
\end{align}

   Its interaction energy becomes:
\begin{align}
	E_\text{int}=\frac{UN_s}{2}\left(1+\frac{1}{2}\cos^2\theta\right)
\label{int12}
\end{align}
   whose minimization leads to $  \theta_\text{min}=\theta_0= \pi/2 $.
   Indeed,  setting $ c_3=c_4=0 $ in Eq.\ref{QQ} leads to $ Q= \bar{c}_1 c_2+ i c_1 \bar{c}_2 =0 $ which, in turn,
   also leads to $ \theta_0= \pi/2 $.

   The order from quantum disorder analysis in Eq.\ref{k1k2} selects $ \phi=0 $ ( Fig.\ref{k1k3fig}a )
   which corresponds to the plane wave state $ c_1=1, c_2=0 $ or $ c_1=0, c_2=1 $.

    Expanding $ E_\text{int} $ around the $ \theta_0=\pi/2 $ and
    expanding the quantum fluctuations corrected quantum ground state energy around the saddle point at $ \phi=0 $
    in the Fig.\ref{k1k3fig}a leads to:
\begin{align}
    H^{(2)}_{12}=E_0+\frac{1}{2}A_2(\delta\theta)^2+\frac{1}{2}B_2( \delta\phi )^2
\end{align}
    where $ A_2=U n_0^2/2 $ is from the classical contribution in Eq.\ref{int12} and $ B_2 \sim (n_0 U)^2/t_s $
    is from the quantum fluctuations which can be extracted  from  Fig.\ref{k1k3fig}a.

  The commutation $ [ \frac{1}{2}n_0 \delta\theta, \delta\phi ]= i \hbar $ leads to the  roton gap
  at $ ( \pi,0) $:
\begin{align}
	\Delta_{2R}=2\sqrt{A_2B_2}/n_0=\sqrt{2UB_2} \sim n_0U \sqrt{ U/t_s}
\label{R2}
\end{align}
  whose numerical values are shown in Fig.\ref{roton2gap}.
  The $ [C_4 \times C_4 ]_D $ symmetry dictates  $ \Delta_{2R}= \Delta_{4R} $.

{\sl (b). Calculations of $\Delta_{3R} $}

The Spin-orbital rotation $R_z^{soc}( \phi )=e^{i \phi \sum_i (-1)^{i_x+i_y}\sigma_z}$ acting on the $ \chi_1 $ state
generates a superposition of condensations at $\mathbf{K}_1$ and $\mathbf{K}_3$.
 The most general state in such a superposition is:
\begin{align}
    \Psi^{0}_{13}=c_1\chi_1e^{-i\mathbf{K}_1\cdot\mathbf{r}_i}+c_3\chi_3e^{-i\mathbf{K}_3\cdot\mathbf{r}_i}
\label{k1k3}
\end{align}
  where $ \chi_1, \chi_3 $ are ortho-normal as listed in Eq.\ref{chi1234}.

  In the following, we will work in this sub-manifold to calculate the gap $\Delta_{3R} $.
  Of course, due to the $ [C_4 \times C_4 ]_D $ symmetry, $\mathbf{K}_2$ and $\mathbf{K}_4$ work equally well
  with   $  \Psi^{0}_{24} $ in Eq.\ref{k1k3}.


   Using $ \chi_1 $ and $ \chi_3 $, we can construct the two eigenstates of $ Q_z= \sum_i (-1)^{i_x+i_y}\sigma_z $ as
\begin{align}
    \Psi_{\text{Z},\pm}
	=\frac{1}{\sqrt{2}}
	\left[\chi_1 e^{-i\mathbf{K}_1\cdot\mathbf{r}_i}
	\pm  \chi_3 e^{-i\mathbf{K}_3\cdot\mathbf{r}_i} \right]
\end{align}
 which can be used to re-parameterize Eq.\ref{k1k3} as:
\begin{align}
    \Psi_i=e^{i\phi/2}\cos(\theta/2)\Psi_{\text{Z},+}+e^{-i\phi/2}\sin(\theta/2)\Psi_{\text{Z,-}}
\end{align}
where
\begin{align}
    c_1 & =[e^{i\phi/2}\cos(\theta/2)+e^{-i\phi/2}\sin(\theta/2)]/\sqrt{2},\quad     \nonumber  \\
    c_3 & =[e^{i\phi/2}\cos(\theta/2)-e^{-i\phi/2}\sin(\theta/2)]/\sqrt{2}
\label{eq:c1+c3}
\end{align}

   Its interaction energy is independent of both $ \theta $ and $ \phi $.
   Indeed,  setting $ c_2=c_4=0 $ in Eq.\ref{QQ} automatically leads to $ Q =0 $ which is independent of
   $ c_1 $ and $ c_3 $.

   The order from quantum disorder analysis in Eq.\ref{k1k3}
   selects $ \theta=\pi/2 $ and $ \phi= \pm \pi/2 $ ( Fig.\ref{k1k3fig} )
   which corresponds to $ (c_1,c_2,c_3,c_4)=1/\sqrt{2} (1,0, \pm i,0) $
   leading to Eq.\ref{k1k30}
   (or  corresponding to  $ (c_1,c_2,c_3,c_4)=1/\sqrt{2} (0,1,0, \pm i ) $
   for $\mathbf{K}_2$ and $\mathbf{K}_4$ in Eq.\ref{k1k3} leading to Eq.\ref{k2k40}. )

   Expanding the quantum ground state energy around the minimum  $ \theta=\pi/2, \phi=\pm \pi/2 $
   in the Fig.\ref{k1k3fig} leads to:
\begin{align}
    H^{(2)}_{13}=E_0+\frac{1}{2}A_3(\delta\theta)^2+\frac{1}{2}B_3( \delta\phi )^2
\end{align}
    where $ A_3 \sim (n_0 U)^2/t_s $ and $ B_3\sim (n_0 U)^2/t_s $ can be extracted
    from Fig.\ref{k1k3fig}a and b respectively. Both  come from the quantum fluctuations.
    It is constructive to compare this equation with Eq.\ref{AB}  generating the roton gap in Fig.\ref{roton1gap} on the left side.

  The commutation $ [\frac{1}{2} n_0 \delta\theta, \delta\phi ]= i \hbar $ leads to the  roton gap at $ ( \pi,\pi) $:
\begin{align}
	\Delta_{3R}=2\sqrt{A_3 B_3}/n_0=\sqrt{2UB_2} \sim n_0 U ( U/t_s )
\label{R3}
\end{align}
  which is shown in Fig.\ref{roton2gap} and smaller than $ \Delta_{2R} $ by a factor $ \sqrt{U/t_s} $.
  Of course, the first order transition as $ h \rightarrow 0 $ is driven by the roton dropping at $ (\pi,\pi) $ with the gap $ \Delta_{3R} $.
  $ \Delta_{2R} = \Delta_{4R}  $ are essentially irrelevant to the first order transition.

{\sl (c). Most general analysis in the whole degenerate manifold. }

 In order to do a most general  analysis, following similar order from disorder analysis
 on the left side of Fig.1, we write the spinor field as the  condensation part Eq.\ref{fournodes}
 plus a quantum fluctuating part $ \Psi= \sqrt{N_0} \Psi_0+ \psi $.
 The zeroth order term $ E_0=-\frac{1}{2}U n_0N_0 $  is the classical energy of the condensate.
 The vanishing of the linear term sets the value of the
 chemical potential. Diagonizing  $ \mathcal{H}^{(2)} $ by a generalized $ 16 \times 16 $  Bogliubov transformation leads to:
\begin{align}
    H
	=E_{0t} +\sum^{8}_{n=1,k\in\text{RBZ}}\omega_{n,k}\left(\alpha_{n,k}^\dagger\alpha_{n,k}+\frac{1}{2}\right)
\end{align}
   where $ -\pi/2<k_x,k_y<\pi/2 $ is the RBZ.

   The ground state energy incorporating the contributions from the quantum fluctuations is:
\begin{align}
    E_\text{GS}[c_1,c_2,c_3,c_4]
	=E_{0t}+\frac{1}{2}\sum^{8}_{n=1,k\in\text{RBZ}}\omega_{n,k}
\end{align}


For general 5 dimension classic ground state manifold,
we know 1 dimension is come from the particle number conservation $ U(1)_c $ symmetry broken
and any state differ by a globe phase have the same energy.
We performed a grid search in the whole 4-dimension manifold
and confirmed that equal superposition of $\mathbf{K}_1$ and $\mathbf{K}_3$ ( or $\mathbf{K}_2$ and $\mathbf{K}_4$ )
condensation Eq.\ref{k1k30} (or Eq.\ref{k2k40} ) have the lowest energy,
therefore are the  $ N=4 $ XY-AFM  quantum ground state with the total degeneracy $ d=2+2=4 $.

{\bf 3. The kinetic energy and density current in all the phases in Fig.1 }

   Using the method in \cite{yan}, we will evaluate the conserved density current in all the phases, especially in the $ N=2 $ and
   $ N=4 $ XY-AFM phases. In the context of 2d charge-vortex duality \cite{yan},
   the vortex currents in the dual lattice gives the boson densities in the direct lattice.
   Here, it would be interesting to see if  the two XY-AFM have different current distributions.
   However, as shown in the following, the current turns out to be zero in all the phases.
   This confirms further the $ N=2 $ and $ N=4 $ XY-AFM phase can not be distinguished by any symmetry breaking principles.
   There is a TPT between the two phases.

   We can write the kinetic energy in Eq.1 as
\begin{align}
    \mathcal{H}_\text{hop}
	=\sum_{i,\mu}(\psi_i^\dagger H_\mu \psi_{i+\mu}+h.c.),
\end{align}
   where $ H_{\mu}=-t\sigma_z + it_s\sigma_\mu,  \quad\mu=x,y $.

  Along a given bond $ (i, i+ \mu) $:
\begin{equation}
   Re [ \psi_i^\dagger H_\mu \psi_{i+\mu} ] =K-i I
\end{equation}
  where $ K $ is the kinetic energy and $ I $ is the current flowing along the bond \cite{yan}.

   For a genetic plane-wave state $ \psi_i=e^{-i\mathbf{Q}\cdot\mathbf{r}_i}\chi $,
\begin{align}
	\psi_i^\dagger H_{\mu}\psi_{i+\mu}
	=e^{-i\mathbf{Q}\cdot\boldsymbol{\mu}}
	\chi^\dagger(-t\sigma_z + it_s\sigma_\mu)\chi
\end{align}

   In the following, we evaluate the kinetic energy and the
   conserved density current in all the 4 phases respectively.
   They can be directly measured by the {\sl In Situ } experimental techniques. 


 (a) For the Z$\uparrow$-FM state
\begin{align}
    \psi_i=\chi_0\>,
\end{align}
  One can easily evaluate
\begin{align}
K_x=K_y=-t;~~~~I_x=I_y=0
\end{align}

 (b) For one of the four Z-XY FM state in Eq.\ref{spinorn}
\begin{align}
	\psi_i=e^{-i\mathbf{K}\cdot\mathbf{r}_i}\chi_\mathbf{K},~~~\mathbf{K}=(k_0,k_0)
\end{align}
where
\begin{align}
	\chi_\mathbf{K}
	=\begin{pmatrix}
		-e^{-i\pi/4}\sin(\theta_0/2)\\
		\cos(\theta_0/2)\\
	\end{pmatrix}
\end{align}
   It is easy to evaluate
\begin{align}
	\psi_i^\dagger H_x\psi_{i+x}
	=\psi_i^\dagger H_y\psi_{i+y}
	=e^{-ik_0}(t\cos\theta_0-it_s\sin\theta_0/\sqrt{2})
\end{align}

  Using  the explicit expressions $ k_0 $ and $ \theta_0 $ in Eq.\ref{twonn}, we find
\begin{align}
   K_x & =K_y=-\frac{2t_s^2}{\sqrt{t_s^2\bigg(8+\frac{h^2}{t_s^2-2t^2}\bigg)}} \leq -t   \nonumber  \\
     I_x & =I_y=0
\end{align}
thus there is no current in this case either.

 It is easy to see that at equality holds at the phase boundary between the Z FM and the Z-XY FM in Fig.1

 (c) For the $ N=2 $ XY-AFM at $h=0$ \& $t_s<\sqrt{2}t$

   Taking one of the  four  $ N=2 $ XY-AFM in Eq.\ref{deg}
\begin{align}
    \psi_{Li}=\frac{1}{\sqrt{2}}
	(\chi_0+e^{i\pi/4}e^{i\mathbf{Q}\cdot\mathbf{r}_i}\chi_\pi)
\end{align}
 we have
\begin{align}
  K_x=K_y= & = -t+\frac{1}{\sqrt{2}}(-1)^i t_s,   \nonumber  \\
    & I_x=I_y=0
\label{n2K}
\end{align}
  so there is  no current.

 (d) For the $ N=4 $ XY-AFM at $h=0$ \& $t_s>\sqrt{2}t$.

   Taking one of the two states in Eq.\ref{k1k30}
\begin{align}
    \psi_{Ri}=\frac{1}{\sqrt{2}}
	(e^{-i\mathbf{K}_1\cdot\mathbf{r}_i}\chi_1
	+ie^{-i\mathbf{K}_3\cdot\mathbf{r}_i}\chi_3)
\end{align}

we have
\begin{align}
  K_x=K_y= & -\frac{1}{\sqrt{2}}t_s-(-1)^i t   \nonumber  \\
    & I_x=I_y=0
\label{n4K}
\end{align}
  so there is  no current.

  By comparing the kinetic energies in the Eq.\ref{n2K} with Eq.\ref{n4K},
  we can see that the lowest kinetic energy bond vanishes at the TT, then changes its sign.
  This salient feature maybe related to the two crossing flat directions $ k_x=\pm k_y $ shown in Fig.2b.

{\bf Acknowledgements }

  We thank Bao Zhen, Wu Zan for helpful discussions.
  F. Sun and J. Y acknowledge AFOSR FA9550-16-1-0412 for supports.
  The work at KITP was supported by NSF PHY11-25915.
  J.W and Y.D thank the support by NSFC under Grant No. 11625522 and the Ministry of Science and
  Technology of China (under grants 2016YFA0301604).




\begin{thebibliography}{99}
\bibitem{naive}  Naively, one may think the $ N=2 $ BEC condensation has the $ C_4 $ Rotation symmetry, while the $ N=4 $
one does not. However, the symmetry of the Hamiltonian is $ [C_4 \times C_4]_D $ instead of just the  $ C_4 $.
So the 4-fold $ N=2 $ XY-AFM and  4-fold $ N=4 $ XY-AFM transform exactly the same way under the $ [C_4 \times C_4]_D $ transformation.
\bibitem{z2} We did not write the dynamic exponent as $ z=2 $, becasue Eq.\ref{lif} only has a cubic $ [ C_4 \times C_4]_D $ symmetry instead of the emergent rotational symmetry.
\bibitem{rotation11} It can be shown that the bosonic wavefunctions Eq.\ref{k1k30}
 differ from Eq.\ref{deg} with $ \phi=\pi/4, 5 \pi/4 $ by a position dependent $ U(1)_c $ phase and
 a global spin rotation around the symmetry breaking axis along the $ (1,1) $ direction.
 Similarly, the bosonic wavefunctions Eq.\ref{k2k40}
 differ from Eq.\ref{deg} with $ \phi=3\pi/4, 7\pi/4 $ by a position dependent $ U(1)_c $ phase and
 a global spin rotation around the symmetry breaking axis along the $ (1,-1) $ direction.
\end{thebibliography}
\end{document}